\documentclass[12pt,preprint]{aastex}
\raggedright

\newcommand\etal{{\em et\,al.}}                 
\newcommand\eg{{\em e.g.}}                      
\newcommand{\HI}{{\rm H\/{\sc i}}}                        
\newcommand\Halpha{{H$\alpha$}}

\newcommand\kms{\,km\,s${^{-1}}$}
\newcommand\msun{M${_\sun}$}
\newcommand\lsun{L${_\sun}$}
\newcommand\MHoverLB{M$_{\rm HI}$/L$_{\rm B}$}

\newcommand\Gyr{\,Gyr}
\newcommand\mm{\,mm}
\newcommand\cm{\,cm}

\newcommand\pc{\,pc}
\newcommand\kpc{\,kpc}

\newcommand\Mpc{\,Mpc}
\newcommand\Vsys{$V_{sys}$}
\newcommand\Vrot{$V_{rot}$}
\newcommand\keer{$\times$}

\newcommand\pg{{\ifmmode {p_g} \else {$p_g$}\fi}}
\newcommand\ig{{\ifmmode {i_g} \else {$i_g$}\fi}}
\newcommand\nng{{\ifmmode {{\bf n}_g} \else {{\bf n}$_g$}\fi}}
\newcommand\llg{{\ifmmode {{\bf l}_g} \else {{\bf l}$_g$}\fi}}
\newcommand\mmg{{\ifmmode {{\bf m}_g} \else {{\bf m}$_g$}\fi}}
\newcommand\tilt{{\ifmmode {tilt} \else {$tilt$}\fi}}
\newcommand\twist{{\ifmmode {twist} \else {$twist$}\fi}}

\begin{document}

\title{The remarkable warped and twisted gas disk in NGC~3718}

\author{Linda S. Sparke}
\affil{University of Wisconsin,
Department of Astronomy, Madison WI 53706}
\email{sparke@astro.wisc.edu}

\author{Gustaaf van Moorsel} 
\affil{ P.~O.~Box 0, National Radio Astronomy
Observatory, Socorro, NM 87801} 
\email{gvanmoor@nrao.edu}

\author{Ulrich J. Schwarz}
\affil{Department of Astrophysics, Radboud University,
 PO Box 9010, NL-6500 GL Nijmegen and \\
Kapteyn Astronomical Institute, 
PO Box 800, 9700 AV Groningen, the Netherlands} 
\email{ulrichs@astro.ru.nl}

\author{Martin Vogelaar}
\affil{ Kapteyn Astronomical Institute, 
PO Box 800, 9700 AV  Groningen, the Netherlands}
\email{vogelaar@astro.rug.nl}

\begin{abstract}

We have mapped NGC~3718, a nearby bright galaxy in a loose group, and its companion NGC~3729 in the 21\cm\ line of neutral hydrogen.
NGC~3718 is a strikingly unusual galaxy with a strong straight dust lane across the center, peculiar diffuse spiral arms, and an extended disk of neutral hydrogen.  
Earlier work showed the gas disk to be strongly twisted, warping
through edge-on where we see the straight dust lane; stars formed in
this gas appear to make up the `spiral arms'.
Our improved maps show a twisted but bisymmetric disk of gas extending to 7\arcmin\ or 35\kpc, where the orbital period is roughly 1\Gyr.  
It is surrounded by fragmentary spiral features, and a streamer of gas extending to a cloud lying 12\arcmin\ or 60\kpc\ to the north.  
We use {\sc inspector}, a task in {\sc gipsy}, to fit a tilted-ring
model interactively to slices through the \HI\ data cube.  
The apparent major axis swings through 100\arcdeg\ from the innermost gas orbits at 30\arcsec\ radius to the outer edge.  
When viewed in the reference frame of the galaxy's stellar disk, the innermost gas orbits are nearly polar, while the outer rings of gas are tilted at 30\arcdeg--40\arcdeg.
The line of nodes, where the gas orbits pass through the plane of the stellar disk, twists by roughly 90\arcdeg\ about the pole.
We do not see gas orbiting in the plane of the stellar disk.
If we assume that the galaxy's dark halo shares the same midplane, then the observed twist can be explained by differential precession in a dynamical model in which the dark halo is fairly round.  
The run of tilt with radius is close to what is required for the warped gas disk to precess rigidly in the galaxy's gravitational field without changing its shape.
This fact probably accounts for the longevity of the twisted structure.

\end{abstract}


Facilities: \facility{VLA, WIYN, ADS, NED}

\section{Introduction \label{intro}}
\begin{figure}
\plottwo{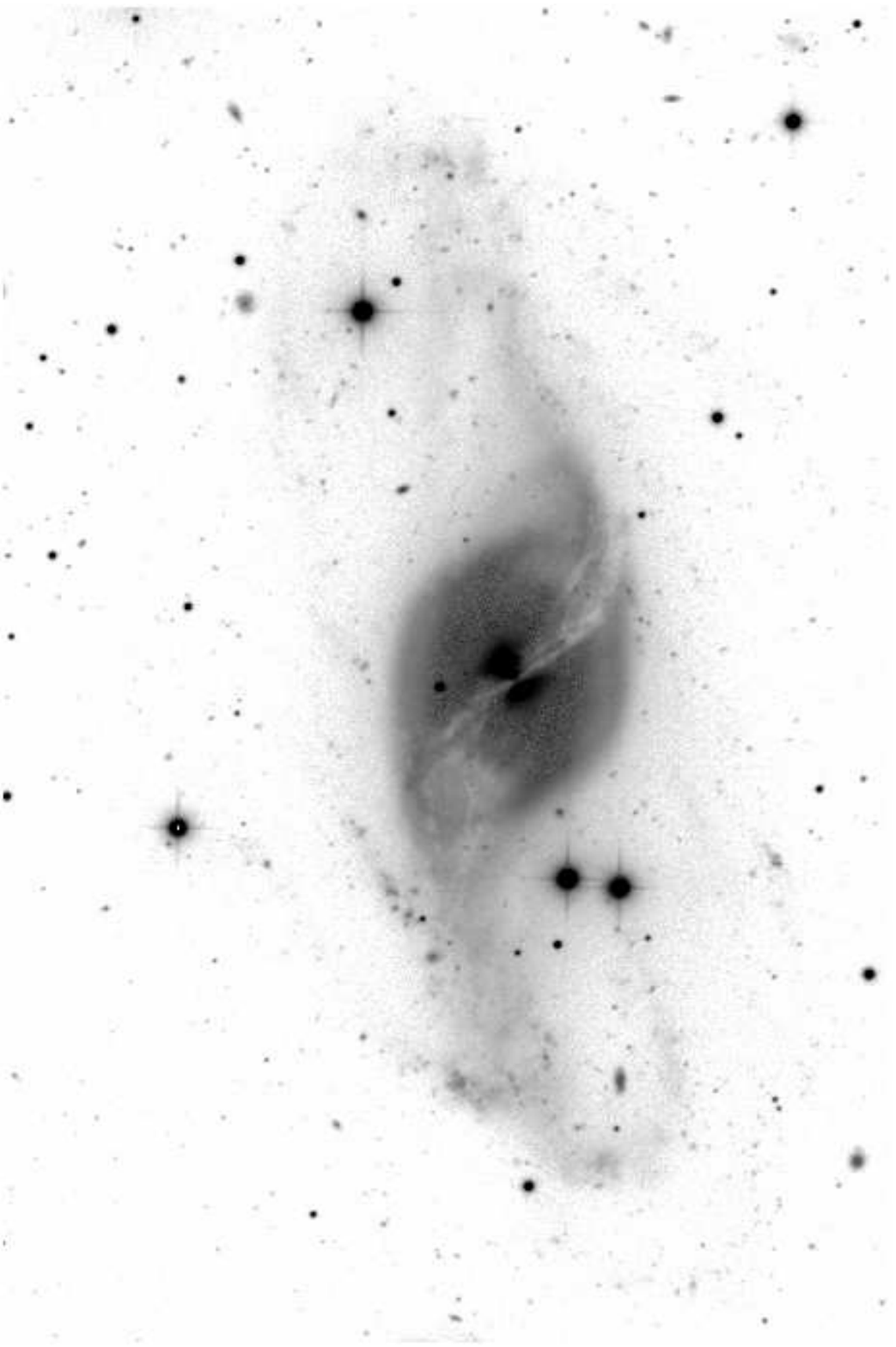}{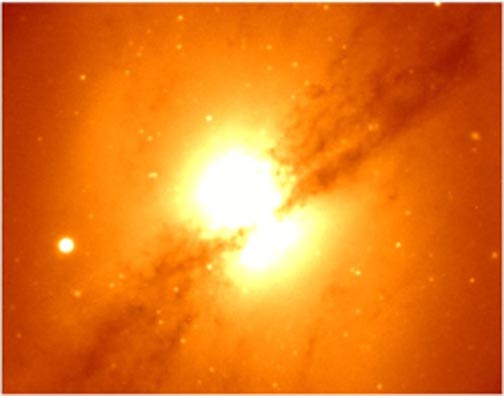}
\caption{Left, B-band image of NGC~3718 taken by E. Wehner
with the WIYN 0.9-m telescope, 
showing the strong central dust lane and diffuse `spiral arms'.
North is up and East is to the left; the image covers 
$9\arcmin \times 13.5\arcmin$. 
Right, $R$-band image of the central region taken by J. Gallagher 
with the 3.5-m WIYN telescope; the dust lane is close to edge-on, with the bright nucleus seen to the north.  The dark feature seen closest to the nucleus is at PA
$\approx125\arcdeg$.}
\label{f_optical}
\end{figure}

The luminous galaxy NGC~3718 (UGC~6524; Arp~214; PRC D-18) 
and its dwarf companion NGC~3729 
form a galaxy pair in the loose Ursa Major group \citep{Tu96}.
Morphologically NGC~3718 is quite peculiar: see Figure~\ref{f_optical}.
A strong dark dust lane resembling that in Centaurus~A 
\citep[NGC~5128; ][]{Du79} 
runs almost edge-on and straight across the central bulge.
Further out, the dust lane diverges into several smooth filaments.
It then twists by almost 90\arcdeg\ into an
`S' shape, forming a diffuse spiral in the stellar light which led 
\cite{RC3} to classify the galaxy as a peculiar barred spiral.
As in Centaurus~A, an active nucleus is largely hidden behind the dust
lane.
A compact radio continuum source less than 0.2\,pc across
has a brightness temperature in excess of $10^7$\,K \citep{Na05};
\cite{Kr07} see a jet stretching 0.5\arcsec\ or 40\pc\ to the north-west.
Optical spectroscopy shows a LINER of Type 1.9 \citep{Ho97} 
with weak broad emission at H$\alpha$ 
and a strong narrower line of [O{\sc i}] at 6300\AA.

The \HI\ maps of \cite{sc85} showed that the dusty gas 
does not lie in the plane of the stellar disk, 
but forms a complex three-dimensional structure.  
Some lines of sight pass more than once through the gas layer, 
giving rise to multiple velocity peaks; 
but the velocity field is strikingly bisymmetric. 
Schwarz was able to describe the gas layer as a violently warped disk made up of material following concentric but tilted orbits, which twisted by roughly 90\arcdeg\ between the inner and outer radii.
The strong straight portion of the dust lane arises where the orbits
turn nearly edge-on to our line of sight, at roughly 200\arcsec\ radius.
\cite{Po04} mapped molecular gas in the inner part of the dust lane
using the CO and HCN lines.
\cite{Kr05} combined those data with interferometric maps
at $\sim 2$\arcsec\ resolution 
in the CO $1 \rightarrow 0$ line at 3\,mm.
They showed that the CO emission
traces an inward extension of the warp that \cite{sc85} derived for the \HI\ gas.
NGC~3718 is included as `related object' D-18 in the Polar Ring Catalog of \cite{PRC}.

Because polar ring systems contain gas orbiting in more than one plane, these rare objects 
constitute one of the few observational probes of the
three-dimensional mass distribution of galaxies.
\citet{Sp90} presented a dynamical model for NGC 3718 to explain the complex shape of the twisted \HI\ disk mapped by \cite{sc85}.  
According to this model we see the underlying disk galaxy almost face-on, with the ring gas in near-polar orbit about it. 
The tilted gas orbits precess about the symmetry axis of 
the flattened central galaxy and its dark halo.  
Orbits at smaller radius precess more rapidly in the galaxy's 
gravitational field, so the gas disk becomes twisted.  
This dynamical model reproduced the main features of Schwarz's tilted-ring fit.
It was consistent with a spherical dark halo, and indicated an age for the gas disk of $3-4$~Gyr.

To test this model, we mapped the system in the 21\cm\ line
of neutral hydrogen with the Very Large Array (VLA) radio telescope.
Our new observations, described in Section~\ref{observ},
improve on those of \cite{sc85} in sensitivity
and in both velocity and spatial resolution. 
In Section~\ref{datacube} we discuss the gas distribution 
and compare our results with previous observations.  
In Section~\ref{tiltmodels} we present tilted-ring fits for the gas motions measured in \HI\ and CO, using optical images to resolve ambiguity about where the gas lies in front of the stellar body.
In Sections~\ref{kinwarp} and \ref{dynwarp} these are interpreted as showing a near-polar disk of gas that has become twisted by the differential precession of the gas orbits in the galaxy's aspherical gravitational potential.

Table~\ref{tablebasic} gives basic information on NGC~3718 and NGC~3729.
We adopt the distance of 17\Mpc\ given by \citet{Tu98} for the Ursa Major group:
there, 1\arcmin\ is equivalent to 4.945\kpc, 1\arcsec\ = 82.4\pc, 
and 13\arcsec\ = 1.07\kpc.
NGC~3718 is a very luminous galaxy with 
$L_B \approx 3 \times 10^{10}$\lsun, while for 
NGC~3729 $L_B \approx 7 \times 10^{9}$\lsun.
\citet{Tu96} classified NGC~3718 as T=1 (Sa) and NGC~3729 as T=2 (Sab) 
on the basis of their optical and near-infrared images.

\section{Observations in the 21cm Line and Data Reduction \label{observ}}

We used the VLA in the C configuration in four
different observing runs in March and April 1992,
for a total observing time of 26 hours and 28 minutes on source. 
To obtain the required velocity coverage and resolution, we observed
using two IFs, tuned at slightly different frequencies in order to
almost double the spectral coverage.  
Parameters of the observations under proposal AS649 are listed in Table~\ref{tableobs}.
Compared to the observations of \citet{sc85}, we improve the velocity 
resolution to 5\kms\ from 33\kms, and the spatial resolution from a
25\arcsec\ $\times$ 31\arcsec\ beam to a 13\arcsec\ circular beam.

The complete data reduction was done using the Astronomical Image
Processing System (AIPS). The four data sets were calibrated
independently, both for amplitude and phase gains errors that vary
with time, and for those that vary with frequency. The absolute flux
scale was determined by observing 3C286, which has a well-known flux
density. After this, the four databases were combined into one.
Inspection of a first mapping of the result allowed us to determine
line-free channels at both edges of the band.  The average of these
channels was subtracted from the uv data set using the AIPS task
UVLIN.  This new data cube, which now contains just line emission, was
used throughout in all subsequent mapping and cleaning.

The continuum map shows point sources at the central locations of both
NGC~3718 and NGC~3729.  
Table~\ref{tablecontinuum}
lists positions and fluxes of these point sources, with uncertainties
0.5\arcsec\ and 1.0~mJy respectively. 
The sources coincide to within this accuracy with the positions given by \citet{VeSa01} at 20\,cm and by \citet{Kr05} at 3\,mm; 
we take them to represent the center of each galaxy.
The presence of a dust lane prevents an accurate optical
position for NGC~3718, but the various values in the literature agree
with this radio position within the error margins.
In their Appendix~B, \citet{VeSa01} quote $11.4 \pm 0.4$~mJy for the
continuum source in NGC~3718, in reasonable agreement with our value of 14.4~mJy; we find 7.9~mJy for NGC~3729, while 
they give a higher flux of $18 \pm 0.9$~mJy.

The data were Fourier transformed using the AIPS task IMAGR, using a
robustness parameter of 0.0, resulting in a circular 13\arcsec\
beam.  
A proper choice of robustness \citep{br95,br99}
results in better sensitivity than when using uniform weighting, 
without the strong non-Gaussian beam effects 
normally associated with natural weighting.
We made two data cubes: one with the full 5~\kms\ velocity resolution,
and once applying a smoothing over the frequency axis resulting in 
10~\kms\ velocity resolution.  
The maps with 10~\kms\ resolution showed no additional structure in the outer galaxy, so 
the 5~\kms\ resolution data were used throughout in this paper.
Both cubes were cleaned to a $1~\sigma$ noise level: 
0.39~mJy/beam for the full resolution data and 
0.30~mJy/beam for the frequency-smoothed data, 
as given in Table~\ref{tableobs}.

\section{Neutral Hydrogen Datacube for NGC~3718 and NGC~3729}\label{datacube}

\subsection{Global Results}\label{globalresults}

\begin{figure}
\includegraphics[height=10cm]{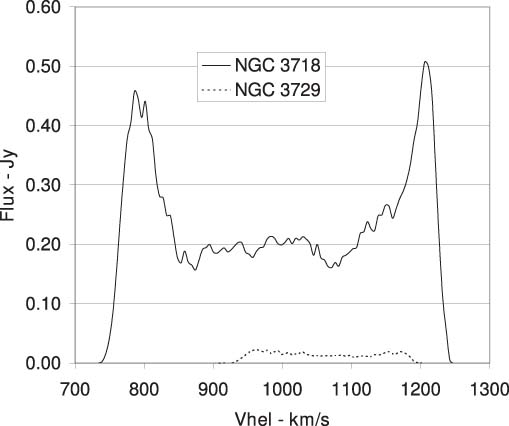}
\caption{Global profiles of \HI\ emission in both NGC~3718 (solid line)
and NGC~3729 (dashed line).
Fluxes have been corrected for primary beam attenuation.}
\label{f_gp}
\end{figure}

We list global \HI\ properties for NGC~3718 and for NGC~3729 
in Table~\ref{tableglobal}.
For both galaxies, we used a method which corrects for the mismatch 
between the dirty beam and the clean beam in the residual map
\citep{jo95} to determine the \HI\ flux in each channel map, 
and corrected for the attenuation of the VLA primary beam. 
The global \HI\ profiles in Figure~\ref{f_gp} show no sign of
absorption against the weak radio continuum source in either galaxy.

For NGC~3718, the \HI\ flux integral of 118~Jy\kms\ agrees with that
found by \citet{sc85} and is 20\% lower than the value of 
\citet{VeSa01}.
Our flux integral of 3.8~Jy~\kms\ for NGC~3729 agrees with the latter authors, but is roughly 5 times lower than given by \citet{sc85}.
A recalculation using the map in Figure~3 of that paper yields a much smaller flux integral, so the result quoted in \citet{sc85} may have been in error.  
Estimates from single-dish observations vary between 90~Jy~\kms\ and
150~Jy~\kms\ \citep{hr89}.
Thus there is little gas in an extended component that would be missed in our maps.

In NGC~3718 we find $8 \times 10^9$\msun\ of \HI\ gas, about twice as
much as in the Milky Way, while the galaxy is about 50\% brighter in
stellar light.  
The ratio \MHoverLB=0.3, which is about average for the
sample of gas-rich S0 and Sa galaxies studied by \citet{No05}.
NGC~3729 has only $3 \times 10^8$\msun\ of \HI\ and is gas-poor
compared to a normal Sab galaxy; we find \MHoverLB=0.04 while 
\MHoverLB=0.1 would be typical \citep{rh94}.

The regular, steep-sided and symmetric profile of NGC~3718 
suggests that the gas has had time to settle into a steady state.
Between the points at which the emission falls to 20\% of its peak
value we measure a width $W_{20}$=476\kms.
If the \HI\ followed pure circular orbits, our measured line width
would yield the rotation speed directly:
$W_{20} = 2 V_{max} \sin i$, where $V_{max}$ is the maximum rotation
speed in the galaxy disk, inclined at angle $i$ to face-on.
\citet{VeSa01} find that in disk galaxies we must subtract about
20\kms\ from $W_{20}$
to correct for random motions in the gas.
For NGC~3718 this would imply $V_{max} \sin i \approx 230$\kms, with
little gas in regions where the circular speed is higher.
 
Based on the mean of the velocities at 20\% of peak flux,
we adopt the systemic velocity \Vsys=995\kms\ for NGC~3718 
and  \Vsys=1063\kms\ for NGC~3729.
For NGC~3718, \citet{VeSa01} derived 993\kms\ from the midpoint of their
\HI\ global profile, and 990\kms\ by examining the position-velocity
diagram along PA=195\arcdeg.  
For NGC~3729 the agreement is even closer: \citet{VeSa01}
find \Vsys=1060\kms\ and 1063\kms\ for the two methods respectively.

\subsection{Channel maps}\label{channelmaps}

\begin{figure}
\includegraphics[height=20cm]{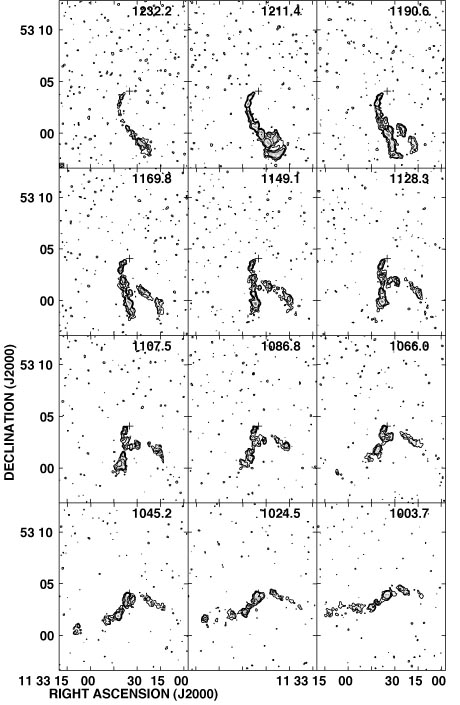}
\caption{Channel maps for the \HI\ distribution in NGC~3718 at
intervals of roughly 20\kms.
Our adopted systemic velocity \Vsys=995\kms\
lies midway between the last channel map
in this figure and the first in Figure~\ref{f_chmap_b}.}
\label{f_chmap_a}
\end{figure}

\begin{figure}
\includegraphics[height=20cm]{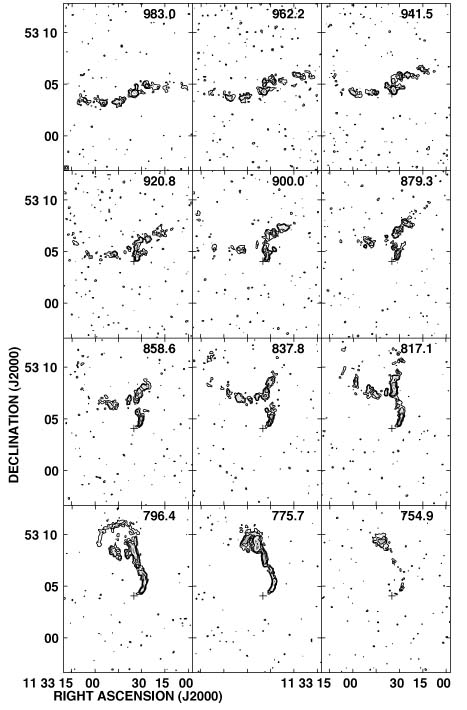}
\caption{Continued from Fig.~\ref{f_chmap_a}: 
channel maps for the \HI\ distribution in NGC~3718.}
\label{f_chmap_b}
\end{figure}

\begin{figure}
\includegraphics[height=15cm]{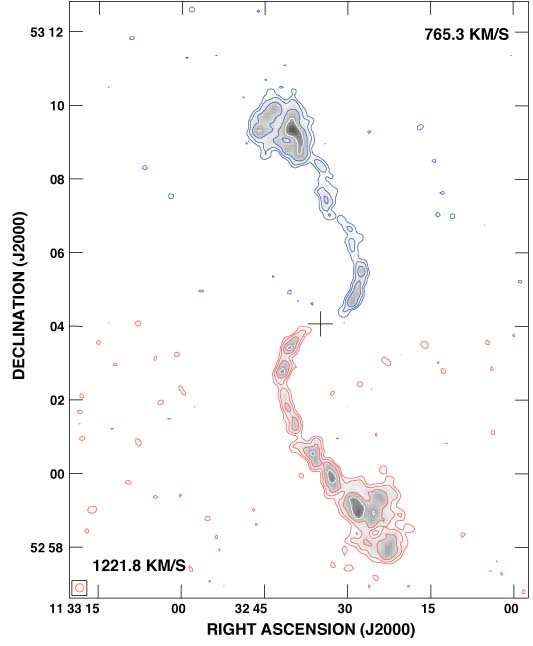}
\caption{Superposed channel maps for the \HI\ gas in
NGC~3718 at velocities displaced roughly 230\kms\ on either side of 
the systemic velocity.
Lower contours, in red, show gas centred at 1221.8\kms;
upper contours, in blue, show gas centred at 765.3\kms.}
\label{extreme_channels}
\end{figure}

Figures~\ref{f_chmap_a} and \ref{f_chmap_b} show the channel maps 
for gas in NGC~3718.
In a warped disk made up of gas on concentric but tilted circular
orbits, gas at each velocity above the systemic velocity \Vsys\
should have a counterpart at the same interval below \Vsys, at a
position point-reflected about the galaxy center. 
In Figure~\ref{extreme_channels}, emission from gas in two extreme
channels centred at 765\kms\ and 1222\kms\ has been superposed to show this symmetry. 

The extreme channel maps containing \HI\ emission are at 755\kms\
and 1232\kms, 
separated by almost exactly our measured width  $W_{20} = 476$\kms.
Channel maps at 5\kms\ lower and higher velocity are empty.
The emission in these channels extends from 30\arcsec\ from the center to 400\arcsec, so 
$V_{rot} \sin i$ should be between 230\kms\ and 240\kms\ 
over this entire radial range.
Within 30\arcsec, either \HI\ gas is largely absent, or 
$V_{rot} \sin i$ is considerably lower.

Within 300\arcsec\ of the center the band of emission in Figure~\ref{extreme_channels} is narrow, suggesting that we see the gas orbits within 10\arcdeg\ to 20\arcdeg\ of edge-on.  
Tracing the ridge line of the emission in the extreme channels then
gives us the position angle of the gas orbits, as plotted in 
Figure~\ref{f_ringangles}.
The kinematic major axis swings from close to PA=100\arcdeg\ at
30\arcsec\ from the center to PA=190\arcdeg\ at radius 400\arcsec.

\subsection{Blanking and moment maps}\label{blank}

The cube of data can be viewed as a rectangular array of velocity (or
frequency) profiles. It is standard practice to reduce the spectral
line data further by forming maps containing the value of the various
moments of each profile. The zeroth moment is a map
showing the spatial distribution of total hydrogen; all velocity
information is lost. In calculating the zeroth moments we restrict
ourselves to that part of the profile where emission is present; this
avoids contamination of the total HI map with noise. Our method of
separating emission and noise is automatic: we convolved the cube to a
40\arcsec~\keer~40\arcsec\ beam, and masked (blanked) all pixels in the {\it high} resolution cube which were below a $3 \sigma$ noise level in the {\it low} resolution cube.  The moment maps are constructed from the unblanked pixels only.
This method avoids the addition of unrelated noise to the total HI map, and at the same time misses little of the low-level HI emission. 

\begin{figure}
\includegraphics[height=18cm]{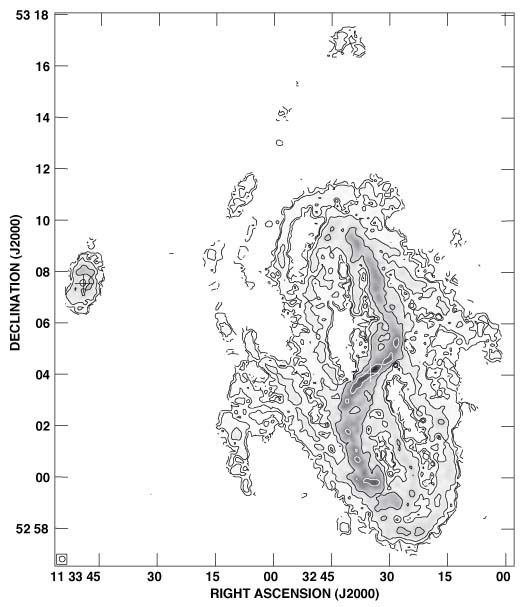}
\caption{Total hydrogen in NGC~3718 and NGC~3729, corrected for the
primary beam attenuation. The lowest contour level is at 
$3.3 \times 10^{19}$~atoms~cm$^{-2}$ or 0.26\msun~pc$^{-2}$, 
approximately at the 3-$\sigma$ noise level. 
Higher contours are at 1.0, 3.3, 10.0, and 33.4~$10^{20}$ 
atoms~cm$^{-2}$. 
The positions of the central continuum sources are marked with
crosses; the beam size is indicated by the small circle in the lower
left hand corner.} 
\label{f_th}
\end{figure}

Figure~\ref{f_th} shows the resulting map.  
Within 100\arcsec\ of the center, we see a narrow dense ridge of \HI\
emission along PA=140\arcdeg, 
coinciding with the dark dust lane of Figure~\ref{f_optical}.  
At larger radii this ridge swings counter-clockwise into a `S' shape.
The left panel of Figure~\ref{f_optical} shows that the diffuse 
`spiral arms' that we see in the starlight correspond to regions  where the \HI\ density in Figure~\ref{f_th} rises above 
$3.3 \times 10^{20}$~atoms~cm$^{-2}$ or 2.6\msun~pc$^{-2}$.
This gas density barely reaches the threshold of 3--10\msun~pc$^{-2}$ normally required for widespread star formation in a galaxy disk \citep[\eg][]{sch04}.  
The tilted-ring model developed for the \HI\ layer in Section~\ref{tiltmodels} below, and illustrated in Figure~\ref{f_velfield}, implies that the projected density along the `spiral arms' is increased by warping in the gas layer. 
The true surface density is even further below the normal threshold for star formation.  

The stellar `spiral arms' extend to roughly 250\arcsec, while the pattern of \HI\ emission is bisymmetric to about 7\arcmin\ or 35\kpc\ from the center, and the \HI\ disk can be traced to a radius of 500\arcsec\ or 41\kpc.
To the southeast a spiral-arm fragment extends to a long streamer, apparently ending in a gas cloud projected 12\arcmin\ or 59\kpc\ from the center. 
This arm fragment and a symmetrically placed structure to the northeast are also visible in the left panel of Figure~\ref{f_optical} as star-forming regions.
As in NGC~1058 and NGC~6946 \citep{fe98, bo05, pr07}, both coherent spiral patterns in the gas and continuing star formation are present far beyond the radius where gravitational instability should be strong enough to initiate them.

Smoothing our data cube at 10~\kms\ velocity resolution further to a 40\arcsec\ beam yields a noise level of 0.5mJy/beam.  A map of total \HI\  made from this smoothed cube fails to show emission more extended than Figure~\ref{f_th}, to a surface density of 0.1\msun~pc$^{-2}$ or $1.3 \times 10^{19}$~atoms~cm$^{-2}$.  In particular, there is no bridge of emission linking NGC~3718 with NGC~3729.

\begin{figure}
\includegraphics[height=18cm]{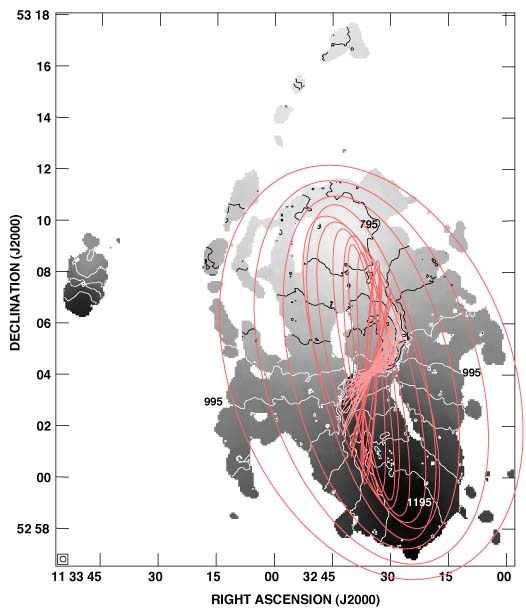} 
\caption{Mean velocity field (first moment map) of the \HI\ emission 
in NGC~3718 and NGC~3729; the tilted ring model of Section~\ref{tiltmodels} for the warped
gas layer of NGC~3718 is superposed. 
Contours are spaced at intervals of 50\kms\ around the systemic
velocity of 995\kms.  All the contours in the gas streamer on the
northeast side of the disk of NGC~3718 are at 795\kms.
Beyond about 200\arcsec\ 
from the center, where velocity profiles are singly
peaked, the values in this map are representative of the radial
velocity of the gas at that position.
The beam size is indicated by the small circle in the lower
left hand corner.} 
\label{f_velfield}
\end{figure}

Beyond 200\arcsec\ from the center, we see only a single velocity peak along each line of sight, and the velocity dispersion is generally below 10\kms. 
Here the first-moment map of Figure~\ref{f_velfield} describes the velocity field of the gas.  
It shows a pattern characteristic of a warped rotating disk: the kinematic major axis (where the velocity is furthest from systemic) twists with radius into an `S'-shape.
The orderly rotation and low velocity dispersion suggests that the structure is at least a few orbits old.   
Beyond about 7\arcmin\ or 35\kpc, the gas of the spiral-arm fragments seen on both sides of the disk along PA=120\arcdeg\ appears to share in the rotation, although it does not form part of a complete ring.
The long streamer curving northwards away from the east side of the
disk is continuous in both position and velocity.
Taking the maximum radius as 500\arcsec\ and the circular speed there as 220\kms\ (see below)  yields a dynamical mass  
M$_{dyn} = 5 \times 10^{11}$\msun, so that M$_{dyn}$/L$_K \approx 7$ in solar units.
This is much larger than the value of unity that is typical of an old
stellar population \citep{be03}; 
so the galaxy must contain substantial dark matter.

The small galaxy NGC~3729 is projected 11\arcmin\ to the east, 
and shows a clear signature of rotation 
in gas that extends to 1\arcmin\ or 5\kpc\ radius.
>From their K-band images, \cite{Tu96} find an isophotal ellipticity 
$e = 1-b/a = 0.32$.  
If we assume the disk to be round (although the galaxy is
classified as barred), it is inclined 48\arcdeg\ from face-on.
Further assuming that the \HI\ gas shares this plane yields a
dynamical mass M$_{dyn} = 35 \times 10^9$\msun\ and M/L$_K \approx 2$ in solar units.
Here we cannot draw strong conclusions about the presence of dark matter.

\begin{figure}
\includegraphics[height=16cm]{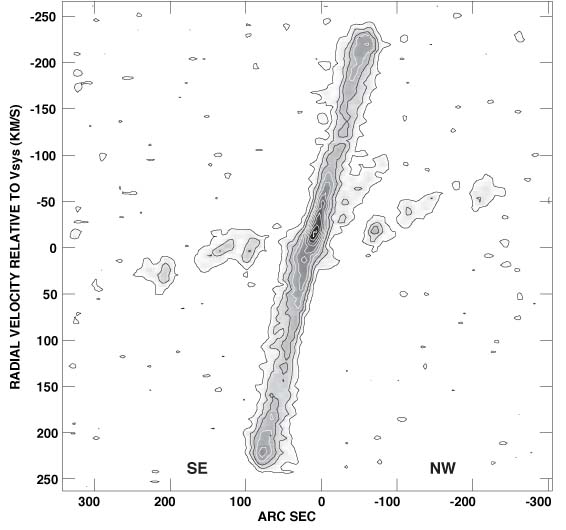} 
\caption{Position-velocity plot through the center of NGC~3718, and
along the apparent HI ridge at PA=140\arcdeg. 
The axes are labeled 
relative to the systemic velocity \Vsys=995\kms\
and the center of NGC~3718: right ascension increases to the left.  
The lowest contour is at 0.75 mJy/beam; 
higher contours are odd multiples.
The sense of the velocity axis is chosen for comparison with the figures of \cite{Kr05}.
}
\label{f_xv}
\end{figure}

In the central parts of NGC~3718 the velocity profiles are neither
singly-peaked nor symmetric.
Figure~\ref{f_xv} shows a position-velocity plot through the center at a position angle of 140\arcdeg,
along the ridge of bright HI emission visible in Figure~\ref{f_th}.
This plot is highly symmetrical, as expected for gas in circular orbit about the galaxy center. 
There are two main components:
the very strong inner one shows velocities rising steeply to 230\kms\ 
at 80\arcsec\ from the center, 
while in the outer component rotation speeds increase almost linearly to 60\kms\ at 300\arcsec\ radius. 
We interpret the slower-rotating gas as following an orbit at larger
radius; we see only the portion projected close to the center, 
where the radial velocity is small.
On the western side and at negative velocities, 
there is a third and much weaker component with an intermediate slope. 
Looking towards any point along this line within 80\arcsec\ of the
center, we would see a double or triple peak in the \HI\ velocity profile.
If the \HI\ gas forms a continuously warped disk 
and we look once through it in the outer parts, 
then each line of sight must cross the gas layer an odd number of
times, so we expect triple profiles.
There is a slight indication that the weak third component may have a
counterpart to the east at positive velocities. 

\cite{Kr05} made interferometric maps of the molecular gas associated
with the inner part of the dust lane.  
They combined several pointings with the single-dish observations of 
\cite{Po04} to probe the dust lane to 70\arcsec\
radius with $\sim 2$\arcsec\ resolution 
in the CO $1 \rightarrow 0$ line at 3\mm.
Their Figure~17 displays a position-velocity diagram along an axis at
PA=130\arcdeg.
Like our Figure~\ref{f_xv}, it is highly symmetric about the center,
with emission at velocities rising to 220\kms\ at 70\arcsec\ radius,
and an even smaller-scale structure where speeds reach 250\kms\ 
within 10\arcsec\ of the center.
This nuclear component would correspond 
to an edge-on disk of diameter 1.5~kpc.
If atomic and molecular gas share the same kinematics, then \HI\ must
be largely absent within 30\arcsec\ of the center; 
otherwise, our Figure~\ref{extreme_channels} would not show a gap in high-velocity emission close to the center.
\citet{rc94} measured velocities in the \Halpha\ line of ionized gas
in the central regions of NGC~3718.  Along PA=130\arcdeg, they found
velocities rising to $260 \pm 20$\kms\ within 80\arcsec\ of the
center, which is consistent with the results in CO.

\section{Tilted-ring models for the \HI\ gas}\label{tiltmodels}
\subsection{Fitting a tilted-ring model}

Because of the very high degree of symmetry in the channel maps of
Figures~\ref{f_chmap_a} and \ref{f_chmap_b}, we follow \cite{sc85}
in modeling the gas within a radius of 500\arcsec\ 
as a strongly-warped but otherwise symmetric disk.
The disk is made up of rings of material, following near-circular
orbits that are concentric but tilted.
Because the emission does not peak symmetrically about a mean velocity
at each point on the sky, we cannot use tasks such as {\sc rotcur}
\citep{Be87, Be89} to determine the ring orientations by fitting to the
mean velocity field, as measured by the first-moment map. 
Instead, we built the task {\sc inspector} in {\sc gipsy} 
\citep{vt01} to compare
the predictions of such a model to various two-dimensional cuts
through the three-dimensional cube of data.

Following the convention of \cite{Ro75} and \cite{Be89}, we measure
the position angle $p$ of each gas orbit anti-clockwise from north to
the line of nodes (the kinematic major axis) on the receding side of
the galaxy.   
Note that this definition of $p$ is 180\arcdeg\ different from that of
\cite{sc85}.  
The orbital inclination $i$ runs from zero as the spin axis points
towards the observer, through 90\arcdeg\ for an edge-on ring, to
180\arcdeg.

Neglecting the effect of both random motion in the gas and our finite
beam size, {\sc inspector} calculates the expected velocities at which
each ring of \HI\ should contribute to a given longitude-velocity cut,
or the positions at which its emission should appear in a given
channel map.
Fixing the central velocity at 990\kms\ gave a slightly better fit than the central value of 995\kms\ that we derived from the global profile.
We placed the ring centers at the radio continuum source, and adjusted the rotation speeds and the ring angles interactively, 
using {\sc inspector} to compare model predictions with the
position-velocity cuts and channel maps.

\begin{figure}
\includegraphics[height=16cm]{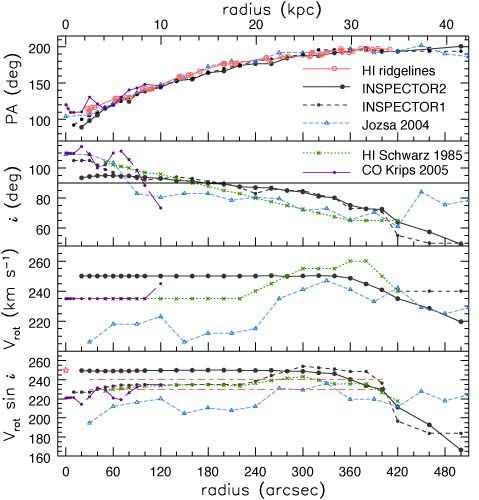} 
\caption{Tilted ring model for the warped \HI\ disk in NGC 3718. 
In the top panel, red lines and open circles show the position angle $p$ of the receding
line of nodes, as defined by the ridge lines of intensity along the three extreme channel maps above and below the systemic velocity (at 754.9\kms, 760.1\kms, 770.5\kms,
1232.2\kms, 1227.0\kms\ and 1216.6\kms).  
Filled circles and a dashed or solid line show our tilted-ring models {\sc inspector}1 and {\sc inspector}2 respectively.
The crosses and green short-dashed line shows the tilted-ring fit derived by \citet{sc85}, while the triangles and the blue (dotted) line show the fit from Figure~6 of \cite{Jo04}.
The purple long-dashed line shows the warped disk of CO from \citet{Kr05}.  
The second panel shows inclination $i$, and the third panel the assumed or fitted circular speed \Vrot. 
The bottom panel gives $V_{rot} \sin i$, the maximum speed along the line of sight. 
The red lines with long and short dashes show the constraint derived in Section~\ref{channelmaps} above
above, and the red star shows the innermost velocities seen in CO by \citet{Kr05}.}
\label{f_ringangles}
\end{figure}

Figure~\ref{f_ringangles} shows that the position angle of the gas orbits is the best-determined quantity.  
In the top panel, we see that at radii $r>40\arcsec$ the results from
{\sc inspector} are in excellent agreement with those obtained by
tracing the ridge line of emission in the extreme velocity channels.
They match fairly well to the model fit by \cite{Kr05} to the CO observations: see below.
Near the galaxy center, the apparent major axis of the gas orbits lies
roughly east-west, almost orthogonal to the major axis of the stellar
light.
It then turns counterclockwise towards north-south, twisting quite 
sharply at smaller radii and then more slowly beyond 300\arcsec.

\begin{figure}
\includegraphics[height=16cm]{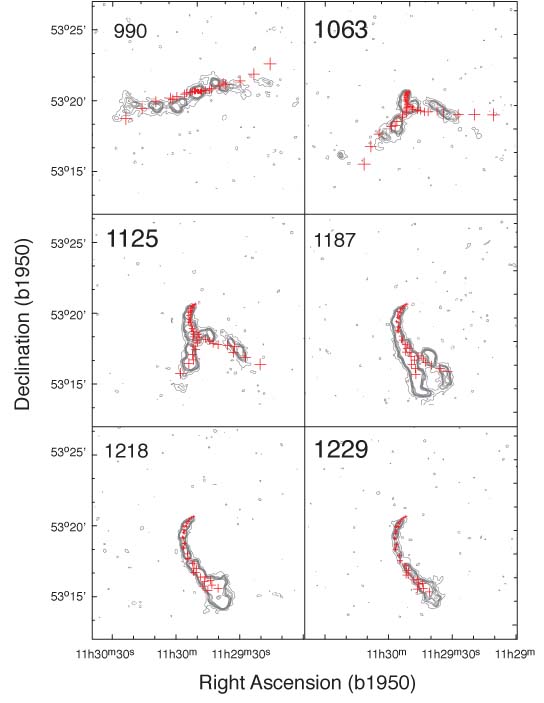}
\caption{Tilted ring model {\sc inspector}2 compared with channel maps at velocities greater than our adopted central velocity of 990\kms. 
Crosses indicate emission from each of the model rings; 
the size of the cross increases proportionally with the ring radius.
The central velocity for each map is given in \kms\ in the top left corner; maps with the larger labels are displaced from the central velocity by the same amount as the maps in corresponding panels of Figure~\ref{f_neg_chmap}.  Other channels are chosen to illustrate features such as the closed contours at 1187\kms.}
\label{f_pos_chmap}
\end{figure}

\begin{figure}
\includegraphics[height=16cm]{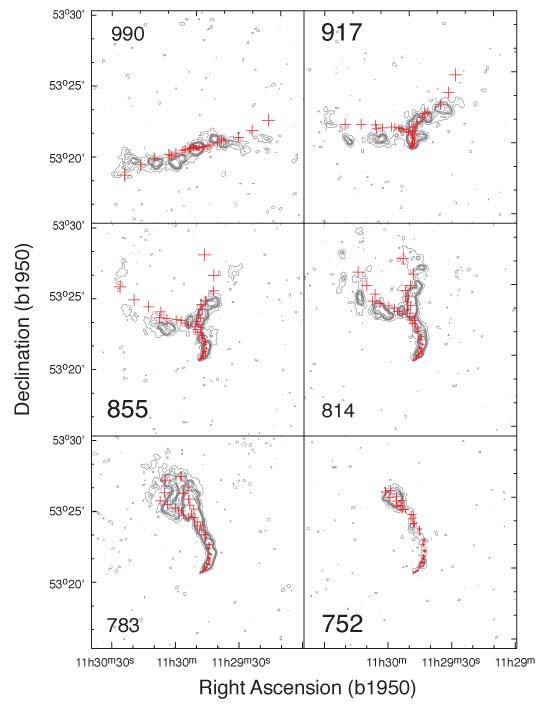} 
\caption{As Figure~\ref{f_pos_chmap}, but for channel maps at velocities below the central velocity.
Maps with the larger velocity labels are displaced from 990\kms\  by the same amount as the maps in corresponding panels of  Figure~\ref{f_pos_chmap}.}
\label{f_neg_chmap}
\end{figure}

\begin{figure}
\includegraphics[height=16cm]{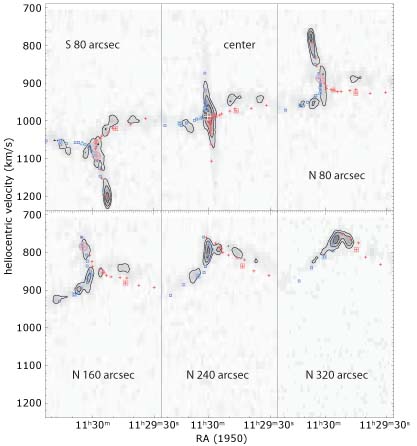} 
\caption{Tilted ring model compared with longitude-velocity cuts
taken along the east-west direction.  
The velocity of the front (closest) portion of each ring is shown by an open blue square, and the rear (more distant) portion by a red cross.
Large squares enclose symbols at radius 400\arcsec; large triangles show rings at 300\arcsec. 
Large red and blue circles with inscribed crosses show rings at 200\arcsec; in the central cut only, these circles almost coincide and are shown as a single light symbol, very close to the central velocity.
Rings at 100\arcsec\ appear in the central cut only, as large circles with central dots.}
\label{f_ring_lvcut}
\end{figure}

The most difficult quantity to determine is the rotation speed.  Initially, we used the rotation curve fit by \cite{sc85} to the earlier \HI\ observations.  
This yielded the model {\sc inspector}1.  However, the CO velocities of \cite{Kr05} show a rise to 250\kms\ within 10\arcsec\ of the center.  
Our 13\arcsec\ synthesized beam is roughly 1\kpc\ across, so we do not expect to resolve a rapid central rise in the rotation curve.  
For the model {\sc inspector}2, we began our iteration with \Vrot\ set at 250\kms, and decreased it in the outer parts only when we could not otherwise obtain a good fit.
Figure~\ref{f_velfield} shows the geometry of this tilted-ring model; the run of position angle and inclination are given in .
Figures~\ref{f_pos_chmap} -- \ref{f_ring_lvcut} compare the model
predictions with position-velocity cuts and channel maps.

The lower panels of Figure~\ref{f_ringangles} show the runs of 
inclination and rotation speed.
The multiply-peaked velocity profiles illustrated in
Figure~\ref{f_xv} require that the warped gas disk passes through
edge-on with $i=90$\arcdeg; we place this transition between 160\arcsec\ and 220\arcsec.
Within this radius, lines of sight can pass three times through the disk.
The position angle in this region of nearly edge-on gas orbits is
155\arcdeg--175\arcdeg, running along the dark central dust lane in the left panel of Figure~\ref{f_optical}.
The velocity field of the \HI\ gas is exactly the same for a ring
of inclination $i$ and one at $180\arcdeg - i$; 
we use the dust distribution in the right panel of Figure~\ref{f_optical} to resolve this ambiguity.
There, we see the bright nucleus to the north of the dust lane; so the south side of the dusty gas disk is closest to us.
The gas recedes on the east side, so its spin axis points away from
us, meaning that $i >90\arcdeg$.
The warp appears smooth as the disk twists through edge-on, 
so we follow \cite{sc85} in assuming that the inclination
decreases monotonically to $i < 90\arcdeg$ at larger radii.

\cite{sc85} constrained the position angle of the edge-on gas orbit by examining the emission peak closest to the center at velocities close to systemic.  He found that the centroid of that peak moved along a line in $PA=-23\arcdeg$ as the velocity decreased through \Vsys.  This is the behavior expected along an edge-on circular orbit in $PA=157\arcdeg$.  We repeated this exercise for the present data set as a consistency check, and find that the central peak moves along $PA=135\arcdeg$. 
Our measured velocity gradient corresponds to a ring at radius 90\arcsec, where
the top panel of Figure~\ref{f_ringangles} shows that the position angle indeed reaches 135\arcdeg.  Thus the gas orbit at this radius is very close to edge-on.

Within 300\arcsec\ of the center the \HI\ gas orbits are less than 10\arcdeg\ from edge-on. 
So the measured speed $V_{rot} \sin i$ should be very near to the orbital speed itself.   
Our stipulation that $V_{rot}$ should decrease monotonically means that our model predicts too high a value for  $V_{rot} \sin i$.  To avoid this we would need  a rotation curve like that of the model {\sc inspector}1, which peaks beyond the optical radius $R_{25}$.
Closed loops in the channel maps at 1190.6\kms\ in Figure~\ref{f_chmap_a} and at 796.4\kms\ in Figure~\ref{f_chmap_b} show that either the rotation speed must drop in the outer disk, or the gas orbits turn closer to face-on.   
We find that both effects are present.
At large radii the shape of the total HI map in Figure~\ref{f_th}
shows that the outermost gas orbits turn to $i \sim 60$\arcdeg, or 
$\sim 30\arcdeg$ from edge-on. They cannot become much more face-on, 
or the predicted east-west extent of gas in channel maps 
near the systemic velocity becomes much larger than observed,
and the two arms of  the ``fork'' in Figures~\ref{f_pos_chmap} and
\ref{f_neg_chmap} are too wide-open.
To reproduce the closed loops, we had to reduce the model rotation
speed to about 220\kms\ near the outer edge.
This behavior is consistent with the 10\%--30\% drop in rotation speed that 
\cite{No07} found to be common in massive S0 and Sa galaxies, with
rotation speeds above 200\kms.   

Far from the galaxy center, emission in the channel map 
near 990\kms\ extends almost east-west for about 200\arcsec\ on each side
of the center.
However, the fit at this velocity (shown in the top left panel in both Figures~\ref{f_pos_chmap} and \ref{f_neg_chmap}) is better on the east (left) side than the west. 
Also, Figure~\ref{f_th} shows that the gas furthest to the east and west does not seem to be part of a complete ring.
So we treat our model with caution within 40\arcsec\ and beyond 400\arcsec\ radius.

We see in Figure~\ref{f_ringangles} that the molecular gas follows the same warped disk structure as the inner \HI\ layer.  
\citet{Kr05} fitted a tilted-ring model to describe the CO
kinematics.
They chose a model rotation curve close to that of \citet{sc85}:
the rotation speed $V(r)$ is taken as 235\kms\ within 100\arcsec\ 
of the center, rising linearly to 255\kms\ at 130\arcsec.
Their Figure~12 displays the derived run of tilt angle with radius, 
relative to a reference plane inclined by 70\arcdeg\ (or 110\arcdeg) 
to the plane of the sky,
and with the {\it approaching} line of nodes at PA=-60\arcdeg.
With respect to that reference plane, their model takes the \twist\ angle to increase with radius $r$ as 
$twist \propto \cos(\tilt) \times  r/V(r)$ (compare Equation~\ref{eqnprec} below).
Taking the reference inclination as $i = 110$\arcdeg\ and 
using the tilt and twist angles kindly supplied by Dr. Krips, 
we recovered the inclination and position angles of their model 
relative to the sky plane, as shown in Figure~\ref{f_ringangles}.
The position angle agrees well with what we derive from the \HI\
observations.  
The inclination oscillates because of the form that they chose for the twist, 
but the product $V_{rot} \sin i$ is very close to that for the \HI\ layer.  

These sets of derived quantities each
represent an eyeball fit to a constrained parametric model.  
This contrasts with systematic fitting techniques such as {\sc rotcur}
that are applied to galaxies with a single-valued velocity field. 
Because the velocity field represents an integral over the full data cube, it is smooth and relatively insensitive to the clumpy distribution of emitting gas, and can be compared directly with a model in which
gas orbits are uniformly filled.  In a galaxy like NGC~3718 we must
work with the full 3-D data cube, where the patchy emitting gas lies
close to  a warped and folded 2-D surface.  

 \cite{Jo04} present a model for the \HI\ layer in NGC~3718 from TiRiFiC, a new method \citep{Jo07} which fits a tilted-ring model automatically to the full data cube.  Their observations at Westerbork had a resolution of 12\arcsec.
Figure~\ref{f_ringangles} shows that the run of position angle is very similar to ours, and the inclination shows the same decreasing trend within 400\arcsec.  
The run of inclination differs; \cite{Jo04} estimate that the gas orbits turn through edge-on closer to the center, at 80\arcsec\ radius and in $PA \approx 130\arcdeg$, and that at larger radii the gas remains further from edge-on than indicated by our model-fits.
In the central 240\arcsec\ the run of $V_{rot} \sin i$ falls below the constraint that we derived from the channel maps of Section~3.2.  
The implied rotation curve is not monotonic, rising from 210\kms\ with maxima at 120\arcsec\ and 320\arcsec. 
The differences between the sets of curves in Figure~\ref{f_ringangles} illustrate the difficulties of the fitting methods, the limitations of the data, and deviations from uniformly filled concentric circular orbits.
 
\subsection{Relation between the gas layer and the stellar disk}

In Figure~\ref{f_optical}, we seem to see the stellar disk of NGC~3718 close to face-on.
We cannot easily measure its orientation from the isophotes, because of the obscuring dust.
From near-infrared photometry in the H band (1.6\,$\mu$m) within 50\arcsec\ of the center,
\cite{PeWi93} find isophotes elongated in PA=112\arcdeg, with
ellipticity $\epsilon \equiv 1 - b/a = 0.17$ (see their Table~4).
\citet{Tu96} give $\epsilon  = 0.58$ at PA=195\arcdeg, measured 
between 150\arcsec\ and 250\arcsec\ radius
(see their Figure~8 and Table~2), 
which would correspond to a round disk seen 55\arcdeg\ to face-on 
(assuming an intrinsic axis ratio $b/a=0.2$). 
In fact the elongation is caused by the stellar light of the `spiral
arms' seen in Figure~\ref{f_optical}. 
Marc Verheijen kindly supplied us with two images at K band  (2.2\,$\mu$m) taken by \cite{Tu96}; only the lower-reslution image with 2.052\arcsec\ pixels was used in their paper.
Neither shows any sign of the dust lane, even at the center. 
>From their higher-resolution image with 0.753\arcsec\ pixels, 
we measure an ellipticity $\epsilon \equiv 1-b/a = 0.11$~to~0.12 
at 25\arcsec -- 27\arcsec, which is 
within the first exponential scale length of the disk but beyond most
of the bulge light (see below).
This would correspond to a round disk with intrinsic $b/a = 0.2$
inclined 28\arcdeg\ to face-on.
The major axis at PA~$\approx 12$\arcdeg\ is almost the same as at
large radii.
However, oval distortions of 10\% are common in galaxy disks \citep{rz95, kk04}, especially among earlier types \citep{ry06}.
In what follows, we assume that the position angle \pg\ where the 
galaxy disk intersects the plane of the sky lies in the range $\pg = 195\arcdeg \pm 20$ \arcdeg.

\citet{HS98} measured velocities along PA=15\arcdeg, 
and find a rise to roughly 100\kms\ at 20\arcsec --30\arcsec\ radius 
on both sides of the center.  
According to their Figure~1 the southwest side of the stellar disk is receding, just as for the outer \HI, so the receding line of nodes lies near PA=195\arcdeg.
If the stellar disk is indeed inclined at $i$=28\arcdeg, 
then for circular speeds close to 250\kms\ we would expect 
to see motions of about 110\kms\ along the kinematic major axis.  
So these observations are consistent with a round stellar disk inclined with its apparent major axis close to their slit position.
\citet{HS98} find a central velocity dispersion of 193\kms\ (including
the factor $f_{bulge}$ in their Table~1), which drops to about 100\kms\
at 30\arcsec.

From this photometric and kinematic evidence, the galaxy disk appears to be nearly face-on, as suggested by the dynamical model developed by \cite{Sp90} for the warped and twisted gas layer.  
However, that model took the plane of the stellar disk to be inclined with $\ig \approx 20\arcdeg$ at a position angle close to PA=--90\arcdeg\ \citep[see][]{Sp02}.
Over most of its radial extent, the \HI\ disk is then tilted by about 80\arcdeg\ with respect to this reference plane, and its twisting could be explained by differential precession.  
But if the stellar disk indeed has this orientation, we would expect streaming speeds to be low along the direction PA=15\arcdeg\ explored by \cite{HS98}.  
It seems more likely that the stellar disk intersects the sky plane along a line closer to PA=15\arcdeg\ (or equivalently PA=195\arcdeg).
Accordingly, we abandon the earlier model.

The gas orbits of our tilted-ring model nowhere lie close to the plane of the galaxy's stellar disk. 
The disk of NGC~3718 seems to be that of an S0 galaxy, substantially free of cool gas.
This is similar to NGC~2655 \citep{Sp08}, an S0/a galaxy with a strong asymmetric central dust lane.
It contrasts with NGC~660 (PRC C-13), a starburst galaxy with a twisted polar ring that is tilted by roughly 55\arcdeg\ to the stellar disk  \citep{vD95}. 
In NGC~660 the host galaxy's disk is as gas-rich as a typical Sc galaxy, and
contains a quarter of the \HI\ gas in the system.

Although it contains $8 \times 10^9$\msun\ of \HI\ gas, with 
$4 \times 10^8$\msun\ of molecular material \citep{Po04}, the modest far-infrared luminosity of $5 \times 10^8$\lsun\ \citep[][with our adopted distance]{Ri88} shows that this galaxy is making few new stars.
The compact star clusters visible in the right panel of Figure~\ref{f_optical} are bluer than their surroundings \citep{Tr07}, indicating relative youth, and the stellar `spiral arms' noted in Section~3.3 show that some starbirth still occurs in this galaxy.
But because the cool gas is not concentrated towards the main stellar disk, its density may be too low for efficient star formation.

\section{Why should the gas layer be warped and twisted?}\label{kinwarp}

What might have caused the gas layer in NGC~3718 to become warped and twisted?  
A disk of material following orbits tilted away from the galaxy equator will tend to twist because of differential precession. 
In an oblate galaxy, consider a cloud of gas following an orbit tipped by an angle 
$\alpha$ away from the equator,
passing upward through the midplane.  
The cloud will make a complete vertical oscillation and again cross the midplane traveling upward, before it has made a whole orbit about the center.  
The tilt of its orbit remains constant, but the line of nodes, where that orbit crosses the symmetry plane, regresses in the direction opposite to the orbital motion: 
\citep[see \eg\ Section~5.8 of][]{Goldstein}.
The angular precession rate $\Omega_p$ for an orbit at radius $r$ inclined by an angle 
$\alpha$ is related to the circular speed $V(r)$ by
\begin{equation}
\Omega_p = { 1 \over {r V(r)} } \,
{ {\partial \langle \Phi \rangle} \over {\partial \cos \alpha} } 
\equiv - \epsilon_{\Phi} \cos \alpha V (r)/r 
\; {\rm or} \;  - \epsilon_{\Phi} \cos \alpha \Omega (r) 
\, .
\label{eqnprec}
\end{equation}
Here $ \langle\Phi \rangle $ represents the gravitational potential
energy, averaged over the ring \citep[\eg][]{Sp86},
and $\Omega (r) = V/r$ is the orbital angular speed.
The quantity $\epsilon_{\Phi} $ measures the flattening of the potential; it is positive for an oblate system, so $\Omega_p$ is negative.
Because the orbital periods are shorter towards the center
the inner orbits will regress faster, 
unless the galaxy's flattening increases strongly with radius.
Thus a gas disk made up of material on concentric tilted orbits generally develops a leading twist.  
Conversely, a disk in a prolate galaxy potential
will twist in a sense that trails the rotation.

Its very regular velocity field suggests that the outer \HI\ disk of NGC~3718 has been in place for at least a couple of orbits. 
For a rotation speed of 230~\kms, the orbital period at 400\arcsec\ is roughly 900\,Myr, implying an
age of at least 2\,Gyr.  
Rotation times in the inner disk are much shorter, and at 40\arcsec\
radius this would correspond to at least 20 orbits.
The position angle of the gas orbits has twisted by about 120\arcdeg\
between these radii. 
If that twist represents precession in a system of roughly constant flattening $\epsilon_{\Phi}$, 
then by Equation~\ref{eqnprec} we must have 
$| \Omega_p | < \Omega /60$ for the inner orbits, or 
$\epsilon_{\Phi} \cos \alpha < 1/60$.
At first glance, this implies that the gravitational potential must be 
improbably spherical to prevent the disk from twisting around
itself many times in its $\geq 2$\Gyr\ lifetime.

The twisted dust lane of the peculiar S0 galaxy NGC~4753 presents a similar dynamical problem.
Here \cite{SCKD92} found a gas disk extending to roughly 7 times the radius of the inner edge at 13\arcsec\ (1\kpc\ at an assumed distance of 15.8\Mpc)
that appears to wrap by almost two complete turns around the galaxy
pole.  They argue that the outer disk is at least six orbits old, corresponding to 40 orbits at the inner edge.
Precessional twisting then implies a nearly spherical mass distribution with axis ratio $b/a \geq 0.84$.
The stellar body is flattened with an axis ratio roughly 2:1 ($b/a=0.5$), so these authors conclude that the dark halo must be both round and gravitationally dominant even well within the main stellar body of this luminous 
galaxy.

A different model for the warped disk of dusty gas in Centaurus~A was proposed by \cite{vA82}: this had the great advantage of representing a stable equilibrium state. 
It requires that the galaxy's mass distribution is not axisymmetric, but a triaxial spheroid tumbling about its short axis. 
The potential then supports a family of stable `anomalous orbits' described by \cite{He82}, which make up a warped disk.  
The dusty gas should settle onto these orbits, in a process studied numerically by \cite{HaIk85}, \cite{HaIk88}, \cite{SCD88} and \cite{CoSp96}.
Near the galaxy center, in the core of the gravitational potential, the anomalous orbits circle the long axis.  
The pole tilts with radius, until at large radii the orbits
lie in the `equatorial' plane perpendicular to the short axis,
circling it in the opposite sense to the one in which the figure tumbles.  
The anomalous orbits make up a twisted disk which follows a
{\it restricted} warp: 
orbits at all radii cross a single line of nodes, which at each
instant lies along the intermediate axis of the tumbling triaxial galaxy.

We can understand the anomalous family as a set of orbits that
precess about the short axis of the galaxy at exactly the right rate
to keep up with the tumbing galaxy potential.
Orbits in a triaxial galaxy will precess about the long or short axis 
at an average rate given by Equation~\ref{eqnprec}, 
where $\epsilon_{\Phi}$ is now the average
flattening about that axis \citep[\eg][]{SCD84}. 
When the triaxial figure tumbles about its own short axis at a rate
$\Omega_t$, the stable anomalous orbit family consists of just those orbits that precess at the rate $\Omega_p = \Omega_t$.
For orbits circling the short axis we have $\epsilon_{\Phi} >0$, so
Equation~\ref{eqnprec} requires the orbital motion to be retrograde with $\Omega_t<0$.
If the system is equally aspherical at all radii and $V (r)$  is
constant, then $\Omega_p$ is constant when
\begin{equation}
\cos \alpha \propto r \, .
\label{anomaloustilt}
\end{equation}
As \cite{He82} point out, the anomalous family tilts over to reach the
galaxy's equatorial plane at the radius where the rate $\Omega_p$ of
free precession for an orbit that is only slightly tilted from the equator becomes equal to the tumbling speed $\Omega_t$. 

To test whether the anomalous orbit family can represent the warped \HI\ layer of NGC~3718, we specify the position angle of each gas orbit by
the unit vector {\bf l} along the receding line of nodes, 
where the orbit crosses the plane of the sky.
The spin axis is along {\bf n} which is perpendicular to
{\bf l}, and we take {\bf m} in the plane of the ring to complete the
right-handed set {\bf l, m, n}.
We take Cartesian coordinates $x,y,z$ from the galaxy center, with $z$
pointing towards the observer, $x$ to the east and $y$ to the north.  
In these coordinates, the vectors {\bf l, m, n} are related to the inclination $i$ and position angle $p$ of Section~\ref{tiltmodels} by    
\begin{eqnarray}
{\bf l} = (- \sin p , \cos p, 0) ~, &
{\bf m} = (- \cos i \cos p , - \cos i \sin p , - \sin i) 
\nonumber \\
{\rm ~and~} &
{\bf n} = (- \sin i \cos p , - \sin i \sin p , \cos i) \, .
\end{eqnarray}
The planes defined by two rings with normals along vectors
${\bf n_1, n_2}$ intersect along the line ${\bf n_1 \times  n_2}$.  
A restricted warp is one in which this vector points in the same
direction for all pairs of rings.  

We can specify the equatorial plane of the galaxy by the apparent position angle \pg\ and inclination \ig\ of a circular ring lying in that plane.
(Note that if the stellar body is triaxial, the position angle of the galaxy's apparent major axis may differ from \pg.)
Defining the corresponding vectors \llg, \mmg\ and \nng,
the angle \tilt\ between a gas ring and that plane is given by 
\begin{equation}
\cos(\tilt) = {\bf n} \cdot \nng = 
\cos i \cos \ig + \sin i \sin \ig \cos(p - \pg) \, .
\end{equation}
The ring intersects the galaxy's equatorial plane along the direction 
${\bf n} \times \nng$.
We define the \twist\ to be the angle in the equatorial plane between 
${\bf n} \times \nng$ 
and the vector \llg\ where the equator intersects the sky plane; so
\begin{eqnarray}
&  \sin(\tilt) \cos(\twist) =  
\llg \cdot {\bf n} \times \nng = 
[\cos \ig \sin i \cos(p - \pg) - \cos i \sin \ig ] =
 {\bf n} \cdot \mmg \; ,
\nonumber \\
& {\rm and}~ 
 \sin(\tilt) \sin(\twist) =  
\mmg \cdot {\bf n} \times \nng  = 
\sin i \sin (p - \pg) =
 -  {\bf n} \cdot \llg \; .
\end{eqnarray}
With these definitions, $\tilt = i$ and $\twist = p-\pg$ when the galaxy's 
equatorial plane coincides with the plane of the sky so that  
$\ig=0$.  
These are related to the angles $\theta, \beta$ of \cite{sc85}
by $tilt = \theta$ and $twist = 180\arcdeg - \beta$.
The pair of angles (\tilt, \twist) describes the same ring as 
($-\tilt, \twist + 180\arcdeg$).
Just as for the inclination $i$, we usually take $0 < \tilt < 180\arcdeg$ so that $\sin(\tilt)$ is positive.  

\begin{figure}
\includegraphics[width=15cm]{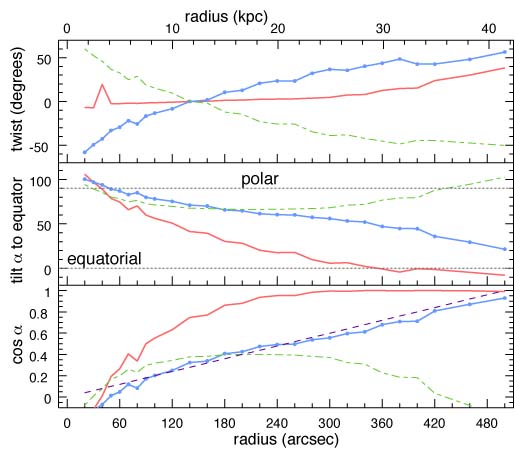}
\caption{For the ring model {\sc inspector}2, angles with respect to a reference plane at inclination \ig\ and position angle \pg.
The top panel shows \twist\ measured relative to the ring at 140\arcsec.  
The red solid line refers to the `restricted warp' obtained for \ig=95\arcdeg\ and \pg=105\arcdeg: the \twist\ is nearly constant in the range 100\arcsec~$< r <$ 400\arcsec.
The blue line with dots refers to the orientation of the stellar disk implied by the K-band isophotes: $\ig$=28\arcdeg, $\pg$=195\arcdeg; the twist is leading.
The green dash-dotted line is for $\ig$=152\arcdeg\ and $\pg$=195\arcdeg, the other possible orientation corresponding to the K-band isophotes. 
The twist now has a trailing sense.
The middle panel shows the angle $\alpha$ between each ring and the galaxy equator in the corresponding model: $\alpha = 90\arcdeg - \tilt$ for the restricted warp and $\alpha = \tilt$ for the other models.  
The ring inclination decreases monotonically with radius from polar to equatorial for all the models except the last.
The bottom panel shows $\cos \alpha$; the dashed line shows the relation $\cos \alpha \propto r$ of Equation~\ref{anomaloustilt}.
The run of \twist\ and \tilt\ for the `restricted warp' model and that with $\ig$=28\arcdeg, $\pg$=195\arcdeg\ is given in Table~\ref{tabletilttwist}.}
\label{f_twotwist}
\end{figure}

The solid curves of Figure~\ref{f_twotwist} show the angles for the \HI\ orbits, relative to a plane with \ig=95\arcdeg\ and \pg=105\arcdeg, 
close to that of the gas ring at $r=40$\arcsec, our innermost reliably-determined orbit.
As \cite{sc85} found, over the best-measured portion of the disk the rings fall close to a restricted warp.
(Schwarz's reference plane corresponds to  $i_g = 104\arcdeg, p_g = 114\arcdeg$ which is close to the position angle of our gas at 50\arcsec\ radius.)
The twist angles of all the rings between 40\arcsec\ and 300\arcsec\ fall within 10\arcdeg\ of a common value.
If we take the ring at 40\arcsec\ to define the polar plane of the galaxy's potential, then 
in $40\arcsec < r < 400\arcsec$ the orientation of the gas orbits changes almost exactly from polar to equatorial, as predicted by the model of \cite{vA82}. 

However, there are two difficulties with interpreting the gas motions as material following anomalous retrograde orbits.  
First, the outer \HI\ orbits should lie perpendicular to the short axis of the triaxial potential.  
Simulations combining dissipative gas with cold dark matter \citep{Du94, Ka04, Ba05} find that this axis tends to be perpendicular to the galaxy disk.  
Those models are directly applicable to luminous galaxies like NGC~3718, where the stellar disk should dominate the gravitational force within the optical radius \citep[\eg][]{Ka06}.
But we saw in Section~\ref{tiltmodels} that the luminous disk is close to face-on, unlike the outer \HI\ orbits.
Also, the shape of the warp does not follow the prediction of
Equation~\ref{anomaloustilt}, given by the straight dashed line in the bottom panel of Figure~\ref{f_twotwist}.
As \cite{sc85} noted, within 200\arcsec\ of the center the tilt of
the disk changes too rapidly to fit this description.  

The stellar disk appears to be tipped by about 28\arcdeg\ from face-on (Section~\ref{tiltmodels} above).  
So we have either $i_g$=28\arcdeg\ or  $i_g$=152\arcdeg, depending on whether the east or the west side of the disk is closer to us.
The blue curves with dots in Figure~\ref{f_twotwist} shows that for the combination
$i_g$=28\arcdeg, $p_g$=195\arcdeg, the central gas disk is very nearly polar while
orbits at larger radius tilt monotonically towards the galaxy plane.  
The bottom panel shows that it follows rather closely the curve $\cos \alpha \propto r$
that we expect for the anomalous orbit family.  
However, this is far from a restricted warp: the twist angle increases
by about 120\arcdeg\ between the inner and outer radii.  
The twist has a leading sense relative to the orbital motion, as we expect for differential precession in an oblate galaxy potential.

When we choose $i_g$=152\arcdeg, the dash-dot line in Figure~\ref{f_twotwist} shows that the tilt is not monotonic.  
The gas orbits are nearly polar near the center, then dip by about 20\arcdeg, and then warp up towards the pole and over it at 420\arcsec. 
Because the flattened stellar body of the galaxy should dominates the gravitational force 
within 150\arcsec\ of the center (see below), the potential should be oblate and we expect the precessional twist to have a leading sense. 
Instead, we see a trailing twist.  
The shape of the gas layer is neither a stable configuration, 
nor a natural result of precessional twisting.
We do not consider this model further, but adopt $i_g$=28\arcdeg\ for the stellar disk.

If the ring is twisted about the galaxy pole, it cannot be in a
steady state: \eg\ \cite{HuTo69}, \cite{Sp86}, \cite{ArSp94}.
Instead, the gas orbits suffer differential precession according to
Equation~\ref{eqnprec}. 
In the following section, we construct a mass model for the galaxy,
to examine how fast the gas layer should twist up, 
and for how long the warped gas disk might have been in place.
This model is similar to those presented for Centaurus A by \cite{Q92}, \cite{Q93} and \cite{Sp96}, 
where the complex warped structure results from an interplay between
self-gravity and precession.

\section{An illustrative dynamical model}\label{dynwarp}

We now examine how a tilted gas disk would precess in a simple axisymmetric mass model for the disk, bulge and dark halo of NGC~3718.  
Our model for the stellar component is based on near-infrared photometry, to minimize the effect of dust absorption.
From deep K-band images that trace the galaxy's light beyond 300\arcsec\ from the center,
\citet{Tu96} measure a  scale length $h_R = 56.6\arcsec = 4.66$\kpc.
This is longer than the $h_R = 27\arcsec$ found by \cite{PeWi93} in the H band, and by \cite{Ch02} from 2MASS K-band photometry, but both of these images were much shallower.  
Making our own ellipse fits to the published image of \citet{Tu96} with 2.052\arcsec\ pixels confirms the longer scale length; so for our illustrative model we adopt $h_R = 55\arcsec$.
We follow \citet{Tu96} in taking the stellar disk to have 
an intrinsic axis ratio $b/a = 0.2$, and calculate the forces from this thickened exponential disk as described in \citet{SS90}.

We take the bulge to be spherical.
The K-band radial profile in Figure~8 of \cite{Tu96} appears roughly exponential outside 20\arcsec, as does the 2MASS profile measured by \cite{Ch02}.  
Since our innermost measured \HI\ orbits lie further out,
it does not matter how we distribute the bulge mass within that radius.
For simplicity we model the bulge as a Plummer sphere with core radius $r_P = 10\arcsec$.

The rotation curve of Figure~\ref{f_ringangles} remains nearly flat to
400\arcsec, which is at least four scale lengths of the stellar disk.  
This, and the high mass-to-light ratio M$_{dyn}$/L$_K \sim 7$ (Section~\ref{blank}), 
requires an extended dark halo.
We use the pseudo-isothermal form of \citet{SS90}, parametrized by 
the flattening $\epsilon = 1 - b/a$ of the equidensity contours, 
the core radius $r_H$ and the asympototic circular speed $V_H$.
For a given halo flattening, we set $r_H$ and $V_H$ by requiring that
the combined rotation curve $V(r)$ from the bulge, disk and halo remains
approximately flat.  
To calculate $V(r)$ we use the equatorial rotation curve of
the halo from Equation~4 of \citet{SS90}, 
but we average the inward pull of the exponential disk over a circular
ring at the appropriate tilt angle. 
The halo torque is computed as described in \citet{SS90}.
The torques from the halo and the flattened disk are added to
calculate the precession rate according to Equation~\ref{eqnprec}.
We do not include the HI gas mass in calculating the rotation curve: see below.
Models like that of \cite{Sp96} for the warped disk in Centaurus~A shows that the self-gravity of the warped disk can also affect details of how it resists precessional twisting.  
In this case the disk is very strongly twisted, so this effect is likely to be small, and we do not include it.

\begin{figure} 
\includegraphics[height=15cm]{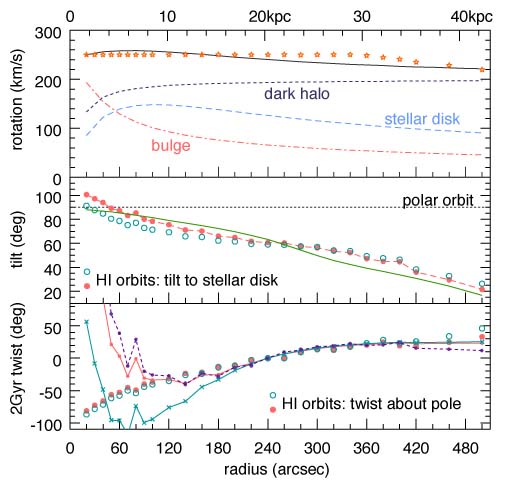}
\caption{Rotation curve and expected twisting for a dynamical model with 
M$_d = 5 \times 10^{10}$\msun\ and M$_b = 2 \times 10^{10}$\msun.
The spherical halo has $r_H$=10\arcsec\ and $V_H$=200\kms.
Top: points show the rotation curve of our tilted-ring fit {\sc inspector2} in Figure~\ref{f_ringangles}; curves show the total rotation predicted from the dynamical model (solid), with the contributions of dark halo (dotted), 
disk (dashed) and bulge (dash-dot).
Middle: angle \tilt\ of the \HI\ orbits from {\sc inspector2} relative to a stellar disk with  \ig=28\arcdeg, \pg=195\arcdeg\ (dashed line with filled dots) and for \pg=175\arcdeg\ (open circles).  
An exactly polar orbit has \tilt=90\arcdeg.
Bottom: angle \twist\ for the \HI\ orbits 
(filled dots for \pg = 195\arcdeg, open circles for \pg=175\arcdeg), and the precessional twisting predicted by the mass model (line with dots for \pg=195\arcdeg, line with crosses for \pg=175\arcdeg). 
The dashed curve shows the result for \pg = 195\arcdeg,
when the halo is flattened with an E3 shape.
All twists are measured relative to the gas orbit at $r=240$\arcsec.
Measured angles and consequently the predicted twists are uncertain
within 40\arcsec\ radius.}
\label{f_prec_fiducial}
\end{figure}

The top panel of Figure~\ref{f_prec_fiducial} shows the rotation curve from this model.
The disk mass M$_d = 5 \times 10^{10}$\msun\ and the halo is chosen to have a small core radius, $r_H$=10\arcsec, so the rotation speed declines gently with radius.
To provide a flat rotation curve at the center, the bulge mass is relatively small, M$_b = 2 \times 10^{10}$\msun;  the overall mass-to-light ratio $M/L_K =1$ in solar units.
This is a `maximim disk' model: 
the disk and bulge must dominate the rotation curve within 2$h_R$ 
to provide the observed declining rotation curve.
We initially take the dark halo to be spherical.  

The middle and bottom panels show the orientation of the gas orbits in our fit {\sc inspector2}, relative to a `galaxy' oriented with $\ig = 28$\arcdeg\ and $\pg = 195$\arcdeg.  
As discussed in Section~\ref{tiltmodels} above, the viewing angles \ig\ and \pg\ for the stellar disk are also uncertain.  
The predicted twist is not very sensitive to a change in the inclination \ig, but decreasing the position angle \pg\ slightly will change the sign of $\cos (tilt)$ for the inner, near-polar orbits, and hence the sense of their precession.  
The open circles show the orientation when we take \pg=175\arcdeg.
  
The bottom panel shows how much twisting a gas disk would suffer over our inferred minimum 2\Gyr\ lifetime, if it was initially warped but not twisted, so that its tilted orbits intersected the `galaxy' plane along a single straight line of nodes.
Comparing this prediction to the twist angles derived from our tilted-ring fit, we see that after 2\Gyr\ the model would develop roughly the observed pattern of twisting in the region between radius 40\arcsec\ and 400\arcsec, where our tilted-ring fit is most reliable.  
Because it warps away from the pole at larger radii, it does not become strongly twisted, as the naive arguments of Section~\ref{kinwarp} would suggest.

The mass of the \HI\ disk is 16\% of that of the stellar disk in our dynamical model.  If we had included it in our rotation curve fit, we would have reduced the mass of both the disk and the dark halo to compensate.  
The torque from the disk would then be no more than 16\% less, and precession times would be longer by that same fraction.  Our conclusion remains unchanged.

Differential precession changes the \twist\ of the gas orbits, but not their \tilt\ with respect to the stellar disk.
Why then should the gas disk have the observed run of tilt?   
In the middle panel of Figure~\ref{f_prec_fiducial} we show the run of tilt angle that allows the orbits at all radii to precess together in our dynamical model.  
The curve is for a halo flattened to an E3 shape, but those for a spherical halo and even an E6 halo lie nearby.
The observed run of tilt lies fairly close to this curve.
We conclude that the warped gas disk has the shape that it does, because that shape has permitted it to survive far longer than would otherwise be the case.  
The dynamical model of Figures~\ref{f_prec_fiducial} assigns the maximum plausible mass to the flattened disk.

\section{Discussion}\label{discuss}

We have made high-resolution maps of the \HI\ gas in NGC~3718 and its companion NGC~3729.  
Our data cube for NGC~3718 shows multiply-peaked velocity profiles and
a complex but highly bisymmetric structure.
Using {\sc inspector}, a task in {\sc gipsy}, we fitted a tilted-ring model, in which gas following near-circular orbits about the galaxy center forms a warped and twisted layer. 
We confirm the conclusions of \cite{sc85}, that the prominent asymmetric dust lane marks the region where the orbits of the (dusty) gas turn edge-on to the line of sight.
The molecular gas mapped by \cite{Kr05} shares the motion of the innermost \HI\ gas.
The unusual diffuse spiral arms fall in regions where gas orbits appear to crowd together on the sky.
The arms are visible in blue light: new stars have formed in the twisted gas layer.
As in other galaxies with extended \HI\ disks \citep{sa08},
spiral structure is observed far out in the disk, where self-gravity should be too weak to provoke the gas to instability and clumping.

The warped and twisted \HI\ disk can be traced to 500\arcsec\ or 42\kpc\ from the center.
It is fairly symmetric within 7\arcmin\ or 35\kpc, where the orbital period is roughly a gigayear.
So the gas disk has probably been in place for at least a few orbits at this radius, or 2--3\Gyr.
Further out, symmetrically-placed spiral-arm fragments to the east and west are visible in both \HI\ gas and blue light.
The polar gas disk is still in the process of formation: the
eastern arm fragment continues as a streamer of gas stretching to a cloud 60\kpc\ north of the galaxy center.  
Sensitive \HI\ maps increasingly reveal such long streamers and tails in the outer parts of disk systems, continuous with the regular velocity field of the galaxy, that may represent gas in the process of joining the galaxy \citep{vdhs05}.
However, the gas in polar orbit around NGC~3718 is very dusty; it is not pristine material.

NGC~3718 has been classified as a barred Sa galaxy, but this is misleading.  
The apparent bar is an effect of looking through the edge-on disk of dusty gas, and the peculiar diffuse spiral arms instead represent star formation in the warped and tilted gas disk.  
K-band photometry \citep{Tu96} shows an exponential disk close to face-on; we find no \HI\ gas orbiting near this plane, so the old stellar disk must be almost empty of cool gas.  
Instead, NGC~3718 is typical of gas-rich early-type galaxies, where \HI\ gas is often found far outside the stellar body, and does not share the stellar kinematics \citep{No05, Mo06, Sp08}.
When we refer our tilted-ring model for the \HI\ gas to the most probable plane for the stellar disk, the innermost gas orbits are nearly polar. 
We do not see gas orbiting in the plane of the stellar disk itself.
Thus NGC~3718 is indeed a polar ring galaxy: as in the archetype NGC~4650A
\citep[\eg][]{G02}, a gas-poor early-type galaxy is surrounded by a highly inclined gas-rich low-surface-brightness disk.

While the inner parts of the \HI\ disk are nearly polar, the outer orbits tip to lower inclination.  
This pattern of tilt minimizes the destructive effects of differential precession, and has allowed the polar structure to survive until the present day.
The observed pattern of twisting can be explained by a dynamical model for the galaxy in which the gas orbits precess freely about the pole of the stellar disk, and the dark halo is roughly spherical.  

Polar ring galaxies are one of our few tools for studying the three-dimensional shape of the dark halo.
Our models for NGC~3718 allows a round dark halo.
\cite{SCKD92} obtain a similar result for the twisted dust disk in NGC~4753, concluding that $b/a>0.8$.  
The Milky Way's flattening can be estimated from the near-polar streams of stars torn from the Sagittarius dwarf galaxy, which undergo differential precession as they orbit our Galaxy.  
This process yields confusing results, with some aspects of the streams pointing to a slightly oblate halo \citep[\eg][]{Jo05} and others to a prolate halo \citep{He04}: see \cite{Fe06} for a summary.  
However, \cite{Jo05} favor the range $0.75 < b/a < 1.1$ for the density, and strongly disfavor a very flattened halo with $b/a < 0.6$.

By contrast, studies of the velocity fields in two polar ring galaxies
imply strongly oblate mass distributions.  
When the dark halo is flattened, polar orbits are generally elongated towards the pole, and the polar rotation curve falls below that in the equatorial plane.
By comparing speeds measured for gas in the polar ring with stellar speeds in the central S0 galaxy, \cite{S94} deduced that the dark halo of NGC~4650A is considerably flattened, with $0.3 \leq b/a \leq 0.4$,
almost as flat as the stellar disk.
In the system A0136-0801, \cite{SP95} used Fabry-Perot imaging to map the two-dimensional velocity field of the polar ring in H$\alpha$ emission.
The kinematic major and minor axes were skewed away from perpendicular, a sign that the gas followed oval orbits.  
Fitting a dynamical model to the velocity field in conjunction with the spatial distribution of the emitting gas yielded a flattening $b/a \approx 0.5$ for the system.

Cosmological simulations predict that the dark halos of galaxies should be fairly round.   
Halos formed from cold dark matter alone should be {\it prolate,} with axis ratios 0.6--0.7: \eg\ \citet{al06}. 
Adding a baryonic disk flattens the halo in the same sense as the disk \citep{Du94},
but only to an average axis ratio 0.7--0.8 \citep{Ka04,Ba05}, 
although \cite{Ka04} found that 'aligned-disk' galaxy mergers could produce a halo as oblate as $b/a \approx 0.5$. 
If the material of the polar ring was a late accretion onto the central galaxy, we would expect the halo to be flattened in the same sense as the host's disk. 
If the ring gas flowed in along filaments of the `cosmic web', as \cite{ma06} propose, the dark halo should be oblate close to the host galaxy, becoming prolate and elongated along the filament further out.  
\cite{bek98} suggested that the polar ring represents the disk of a low-surface-brightness galaxy that captured the dense central body by merger.  
The dark halo might then be aligned with its long axes in the plane of the ring.
Polar rings indeed deviate from the Tully-Fisher relation in the sense that Bekki's model would predict \citep{Io03}; rotation speeds measured from the gas of the polar ring are higher than expected.  

The dark halos of galaxies may really be quite diverse; the review of \cite{Me04} showed measurements covering the whole range $0.2 < b/a < 0.8$.
But why should we see such a pronounced difference among polar ring systems?
Perhaps this is simply observational selection.
Both NGC~4650A and A0136-0801 have `classical' polar rings, lying nearly perpendicular to the host galaxy's stellar disk.
In a galaxy with a flattened halo like NGC~4650A, a gas disk tipped far from the perpendicular as that in NGC~3718 would rapidly become twisted beyond recognition.
Strongly tilted rings would survive preferentially in systems with the roundest halos.  
Another possibility is that the halo shape depends systematically on the galaxy's luminosity. 
The Milky Way, NGC 3718 and NGC 4753 are all luminous systems, while NGC~4650A and A0136-0801 are several times less luminous, with $L_B \approx 4 \times 10^9$\lsun.

\acknowledgments{We are very grateful to Elizabeth Wehner and Jay Gallagher for help with optical and near-infrared images, and especially for Figure~\ref{f_optical}; 
to Melanie Krips for supplying details of her model fit for the molecular gas; 
and to Marc Verheijen for access to his K-band images. 
LSS acknowledges support from the National Science Foundation through grant AST-00-98419.
GvM and LSS would like to thank the Kapteyn Astronomical Institute of Groningen University, Netherlands, 
and the MPI for Astrophysics in Garching, Germany for hospitality while part of this work was carried out.
We are all grateful to Hugo van Woerden for his encouragement throughout this project.
Finally, we would like to thank our anonymous referee for comments that helped us to improve and shorten the paper.

The National Radio Astronomy Observatory is a facility of the National Science Foundation operated under cooperative agreement by Associated Universities, Inc. 
The WIYN Observatory is a joint facility of the University of Wisconsin-Madison, Indiana University, Yale University, and the National Optical Astronomy Observatories.
The NASA/IPAC Extragalactic Database (NED) is operated by the Jet Propulsion Laboratory, California Institute of Technology, under contract with the National Aeronautics and Space Administration (NASA).  
This research has made use of NASA's Astrophysics Data System (ADS).
}

\begin{deluxetable}{lcc} 
\tablecolumns{3} 
\tablewidth{0pc} 
\tablecaption{Basic data for NGC~3718 and NGC~3729 \label{tablebasic}}
\tablehead{\colhead{Parameter} & 
\colhead{NGC~3718}   &  \colhead{NGC~3729}}
\startdata
Type\tablenotemark{a} & SBa(pec) (T=1) & SBa(pec) (T=2) \\ 
Distance\tablenotemark{b}         &  17~Mpc &   17~Mpc     \\            
Corrected apparent B magnitude\tablenotemark{c} &  10.45 & 12.01 \\
B Luminosity & $3 \times 10^{10}$\lsun & $0.7 \times 10^{10}$\lsun \\
Corrected apparent K$'$ magnitude\tablenotemark{c}&  7.35 & 8.57 \\
K$'$ Luminosity & $7 \times 10^{10}$\lsun & $2 \times 10^{10}$\lsun \\ 
Optical radius $R_{25}$\tablenotemark{d} & 3.2\arcmin & 1.3\arcmin \\
Ellipticity $1-b/a$ of stellar disk\tablenotemark{e} & 0.11 & 0.32 \\
Inclination $i$\tablenotemark{f} & 28\arcdeg & 48\arcdeg \\
\enddata 
\tablenotetext{a}{\citet{RC3}; numerical type estimated by \citet{Tu96}}
\tablenotetext{b}{\citet{Tu98}}
\tablenotetext{c}{Extinction-corrected magnitudes $M^{b,i}_B$ and 
$M^{b,i}_{K'}$ from Table 2 of \citet{VeSa01}}
\tablenotetext{d}{From $D_{25}$ in Table~2 of \citet{VeSa01}}
\tablenotetext{e}{For NGC~3718, from Section~\ref{blank}; for
  NGC~3729, from Table~4 of \cite{VeSa01}}
\tablenotetext{f}{We assume a circular disk with intrinsic axis ratio $B/A = 0.2$, so that 
$\cos^2 i = [(b/a)^2 - (B/A)^2]/[1 - (B/A)^2]$}
\end{deluxetable}

\begin{deluxetable}{lc} 
\tablecolumns{2} 
\tablewidth{0pc} 
\tablecaption{Parameters of \HI\ Observations \label{tableobs}}
\tablehead{}
\startdata
Dates of observation                & 1992.25                         \\
R.A. pointing center (J2000.0)       & 11$^h$ 32$^m$ 39.922$^s$            \\
$\delta$ pointing center (J2000.0) & +53\arcdeg\ 5\arcmin\ 25.977\arcsec \\
Synthesized beam                    & $13.5\arcsec \times 13.2\arcsec$\\
Effective Velocity coverage -- \kms &  708 -- 1253                    \\
Number of channels                  &  106                            \\
Velocity resolution -- \kms         &  5.2 / 10.4\tablenotemark{a}    \\
Noise -- mJy/beam (rms)           & 0.39 / 0.30\tablenotemark{a}    \\
Noise -- K (rms)                    & 1.43 / 1.10\tablenotemark{a}    \\
\enddata 
\tablenotetext{a}{For the high- and low- velocity resolution cubes,
  respectively.}
\end{deluxetable}

\begin{deluxetable}{lcc} 
\tablecolumns{3} 
\tablewidth{0pc} 
\tablecaption{20~cm Continuum Emission \label{tablecontinuum}}
\tablehead{\colhead{Parameter} & \colhead{NGC~3718}   &
  \colhead{NGC~3729}}
\startdata
Right Ascension (2000) & 11$^h$ 32$^m$ 34.934$^s$ & %
11$^h$ 33$^m$ 49.356$^s$ \\
Declination $\delta$ (2000) & +53\arcdeg\ 4\arcmin\ 4.63\arcsec &%
+53\arcdeg\ 7\arcmin\ 31.57\arcsec \\
Flux at 20\cm\ (mJy)  &  $14.4  \pm 1$   &   $7.9 \pm 1$   \\
\enddata 
\end{deluxetable}

\begin{deluxetable}{llcc} 
\tablecolumns{4} 
\tablewidth{0pc} 
\tablecaption{Global \HI\ parameters of NGC~3718 and NGC~3729 \label{tableglobal}}
\tablehead{\colhead{Parameter} &  \colhead{Unit} &
\colhead{NGC~3718}   &  \colhead{NGC~3729}}
\startdata
Systemic velocity \Vsys\tablenotemark{a}             & \kms  &  995     &  1063     \\ 
\HI\ full width $W_{20}$\tablenotemark{b} & \kms  &  476     &   242     \\
\HI\ flux integral\tablenotemark{c}      & Jy\kms & 117.7    &   3.8     \\
\HI\ radius 
&  & 8.3\arcmin\ = 41\kpc   &   1.4\arcmin = 5\kpc
\\
\HI\ mass\tablenotemark{d} & $10^9$\msun   &   8.0    &   0.3  \\ 
Dynamical mass\tablenotemark{e}             & $10^9$\msun    &   500 & 35 \\
M(\HI)/L$_B$    &  \msun/\lsun  &   0.3    &   0.07 \\   
\enddata 
\tablenotetext{a}{The velocity symmetric with respect to 20\% of the
peak value on the profile wings.}
\tablenotetext{b}{Full width at 20\% of peak flux as described in the text.}
\tablenotetext{c}{The total area under the global profile.}
\tablenotetext{d}{Calculated from the total flux assuming an optically
  thin medium.}
\tablenotetext{e}{For NGC~3729, adopting the inclination $i$ from Table~\ref{tablebasic}.}
\end{deluxetable}

\begin{deluxetable}{rrrrrrr} 
\tablecolumns{7} 
\tablewidth{0pc} 
\tablecaption{For the model {\sc inspector}2, position angle and inclination derived for the gas orbits, and their tilt and twist angles relative to two reference planes:
(a) that of a `restricted warp' and 
(b) that indicated by the K-band isophotes of the inner galaxy
\tablenotemark{~}\label{tabletilttwist}}
\tablehead{\colhead{Radius} &  \colhead{PA} & \colhead{$i$} &
\colhead{$tilt$(a)}   &  \colhead{$twist$(a)} 
&\colhead{$tilt$(b)}   &  \colhead{$twist$(b)}
\\ \colhead{arcsec} &  \colhead{deg} & \colhead{deg} & 
\colhead{$\ig=95$\arcdeg} & \colhead{$\pg=105$\arcdeg} &
\colhead{$i_g=28$\arcdeg} & \colhead{$\pg=195$\arcdeg} }
\startdata
 20.0  &  89.0 &  93.5 & -16.1 & 85.1 &  100.5 & -102.6 \\
 30.0  &  98.1 &  94.4 &   -6.9 & 84.8 &   97.1 & -94.0 \\
 40.0  & 105.5 &  94.8 &   0.6 & 111.4 &   94.0 & -87.3 \\ 
 50.0  & 116.3 &  95.0 &  11.2 &  89.6 &   89.2 & -77.7 \\ 
 60.0  & 120.5 &  94.9 &  15.4 &  89.6 &   87.2 & -74.0 \\ 
 70.0  & 129.0 &  94.6 &  23.9 &  90.1 &   83.1 & -66.6 \\ 
 80.0  & 124.7 &  94.8 &  19.7 &  89.8 &   85.2 & -70.3 \\ 
 90.0  & 135.1 &  94.1 &  30.0 &  90.5 &   80.1 & -61.2 \\ 
100.0  & 138.8 &  93.7 &  33.7 &  90.8 &   78.3 & -57.9 \\ 
120.0  & 144.3 &  93.1 &  39.2 &  91.3 &   75.5 & -52.9 \\ 
140.0  & 153.4 &  91.7 &  48.4 &  92.1 &   71.1 & -44.5 \\ 
160.0  & 155.2 &  91.4 &  50.3 &  92.3 &   70.2 & -42.8 \\ 
180.0  & 164.3 &  89.7 &  59.5 &  93.3 &   65.9 & -34.0 \\ 
200.0  & 166.6 &  89.2 &  61.8 &  93.6 &   64.8 & -31.7 \\ 
220.0  & 174.2 &  87.5 &  69.5 &  94.5 &   61.6 & -23.8 \\ 
240.0  & 176.8 &  87.0 &  72.2 &  94.8 &   60.5 & -21.0 \\ 
260.0  & 176.7 &  86.7 &  72.1 &  95.1 &   60.2 & -21.1 \\ 
280.0  & 184.6 &  85.0 &  80.1 &  96.0 &   57.5 & -12.3 \\ 
300.0  & 188.5 &  84.0 &  84.1 &  96.6 &   56.2 &  -7.8 \\ 
320.0  & 187.8 &  81.1 &  83.7 &  99.5 &   53.4 &  -8.9 \\ 
340.0  & 191.7 &  80.1 &  87.6 & 100.1 &   52.2 &  -4.2 \\ 
360.0  & 194.5 &  75.2 &  90.8 & 104.8 &   47.2 &  -0.7 \\ 
380.0  & 198.0 &  72.8 &  94.3 & 106.9 &   44.9 &   4.0 \\ 
400.0  & 193.7 &  72.5 &  90.3 & 107.5 &   44.6 &  -1.8 \\ 
420.0  & 193.8 &  64.0 &  91.1 & 116.0 &   36.0 &  -1.8 \\ 
460.0  & 197.1 &  57.5 &  94.4 & 122.3 &   29.5 &   3.6 \\ 
500.0  & 200.8 &  49.3 &  97.6 & 130.5 &   21.6 &  12.0 \\   
\enddata
\end{deluxetable}


\clearpage


\begin{thebibliography}{}

\bibitem[Allgood \etal(2006)]{al06} 
Allgood, B., Flores, R. A., Primack, J. R., Kravtsov, A. V., Wechsler, R. H., Faltenbacher, A., Bullock, J. S. 2006, \mnras, 367, 1781

\bibitem[Arnaboldi \& Sparke(1994)]{ArSp94} 
Arnaboldi, M., Sparke, L. S. 1994, \aj, 107, 958 

\bibitem[Arnaboldi \etal(1997)]{A97} 
Arnaboldi, M., Oosterloo, T., Combes, F., Freeman, K. C.,
Koribalski, B. 1997, \aj, 113, 585 

\bibitem[Bailin \etal(2005)]{Ba05}
Bailin, J., Kawata, D., Gibson, B. K., Steinmetz, M., Navarro, J. F., Brook, C. B., Gill, S. P. D., Ibata, R. A., Knebe, A., Lewis, G. F., Okamoto, T. 2005, \apj, 627, L17. 

\bibitem[Begeman(1987)]{Be87}
Begeman, K. G. 1987, PhD Thesis, Groningen University, Netherlands.

\bibitem[Begeman(1989)]{Be89}
Begeman, K. G. 1989, \aap, 223, 47

\bibitem[Bekki(1998)]{bek98}
Bekki, K. 1998, \apj, 499, 635

\bibitem[{Bell \etal}(2003)]{be03}
Bell, E. F., McIntosh, D. H., Katz, N., Weinberg, M. D.
2003 ApJS 149, 289

\bibitem[Binney \& Tremaine(1987)]{BT87}
Binney J., Tremaine S., 1987, Galactic Dynamics, (Princeton University  
Press)

\bibitem[Boomsma \etal(2008)]{bo05} 
Boomsma, R., Oosterloo, T. A., Fraternali, F., van der Hulst, J. M., Sancisi, R. 2008 \aap, 490, 555


\bibitem[{Briggs}(1995)]{br95}
Briggs, D. S. 1995 PhD thesis, New Mexico Institute of Mining and Technology

\bibitem[{Briggs \etal}(1999)]{br99}
Briggs, D. S, Schwab, F. R., Sramek, R. A. 1999 
ASP Conf. Ser. 180, 127

\bibitem[Chitre \& Jog(2002)]{Ch02}
Chitre, A., Jog, C. J. 2002 \aap, 388, 407

\bibitem[Colley \& Sparke(1996)]{CoSp96}
Colley, W. C., Sparke L. S. 1996, \apj, 471, 748

\bibitem[Combes \& Arnaboldi(1996)]{CoAr96}
Combes F., Arnaboldi M., 1996, \aap, 305, 763

\bibitem[de Vaucouleurs \etal(1991)]{RC3}
de Vaucouleurs, G., de Vaucouleurs, A., Corwin, H. G. \etal, 1991,
Third Reference Catalogue of Bright Galaxies (Springer, New York)

\bibitem[Dubinski(1994)]{Du94}
Dubinski, J. 2004, \apj, 431, 617

\bibitem[Dufour \etal(1979)]{Du79}
Dufour, R. J., Harvel, C. A., Martins, D. M., Schiffer, F. H., III,
Talent, D. L., Wells, D. C., van den Bergh, S., Talbot, R. J., Jr. 
1979, \aj, 84, 284

\bibitem[Fellhauer \etal(2006)]{Fe06}
Fellhauer, M., Belokurov, V., Evans, N. W., Wilkinson, M. I., Zucker, D. B., Gilmore, G., Irwin, M. J., Bramich, D. M., Vidrih, S., Wyse, R. F. G., Beers, T. C., Brinkmann, J. 2006, \apj, 651, 167

\bibitem[Ferguson \etal(1998)]{fe98} 
Ferguson, A. M. N., Wyse, R. F. G., Gallagher, J. S., Hunter, D. A.
1998, \apjl, 506, L19  


\bibitem[Gallagher \etal(2002)]{G02}
Gallagher J. S., Sparke L. S., Matthews L. D., Frattare L. M.,
English J., Kinney A. L., Iodice E., Arnaboldi M. 2002,
\apj, 568, 199 

\bibitem[Goldstein \etal(2002)]{Goldstein}
Goldstein, H., Poole, C., Safko, J. 2002, 
Classical Mechanics (3rd ed.), Addison Wesley (San Francisco), Section 5.8

\bibitem[Habe \& Ikeuchi(1985)]{HaIk85} 
Habe, A., Ikeuchi, S. 1985, \apj, 289, 540

\bibitem[Habe \& Ikeuchi(1988)]{HaIk88}
Habe, A., Ikeuchi, S. 1988, \apj, 326, 84

\bibitem[H\'eraudeau \& Simien(1998)]{HS98}
H\'eraudeau, Ph., Simien, F. 1998, \aaps, 133, 317

\bibitem[Heisler \etal(1982)]{He82}
Heisler, J., Merritt, D., Schwarzschild, M. 1982, \apj, 258, 490

\bibitem[Helmi(2004)]{He04}
Helmi, A. 2004, \apjl, 610, L97

\bibitem[Ho \etal(1997)]{Ho97}
Ho, L. C., Filippenko, A. V., Sargent, W. L. W., Peng, C. Y.
1997, \apjs, 112, 391

\bibitem[Huchtmeier \& Richter(1989)]{hr89}
Huchtmeier, W. K., Richter, O.-G. 1989, `A General Catalog of HI
Observations of Galaxies', Springer-Verlag (New York)

\bibitem[Hunter \& Toomre(1969)]{HuTo69}
Hunter, C., Toomre, A. 1969, \apj, 155, 747

\bibitem[Iodice \etal(2003)]{Io03}
Iodice, E. Arnaboldi, M., Bournaud, F., Combes, F., Sparke, L. S., van Driel, W., Capaccioli, M. 2003, \apj, 585, 730

\bibitem[Johnston \etal(2005)]{Jo05}
Johnston, K.V., Law, D. R., Majewski, S. R. 2005, \apj, 619, 800 

\bibitem[J\"ors\"ater \& Van Moorsel(1995)]{jo95} 
J\"ors\"ater, S., van Moorsel, G.A 1995, \aj, 110, 2037

\bibitem[J\'ozsa \etal(2004)]{Jo04}
J\'ozsa, G., Oosterloo, T., Klein, U., Kenn, F. 2004, in 
`Baryons in Dark Matter Halos', 
eds. R. Dettmar, U. Klein \& P. Salucci, 
SISSA, Trieste, Italy.

\bibitem[J\'ozsa(2006)]{Jo06}
J\'ozsa, G. 2006, PhD thesis, University of Bonn; at
http://hss.ulb.uni-bonn.de/diss\_online/math\_nat\_fak/2006/jozsa\_gyula/index.htm

\bibitem[J\'ozsa \etal(2007)]{Jo07}
J\'ozsa, G. I. G., Kenn, F.,  Klein, U., Oosterloo, T. 2007 \aap, 468, 731


\bibitem[Kassin \etal(2006)]{Ka06}
Kassin, S. A., de Jong, R. S, Weiner, B. J. 2006 \apj, 643, 804

\bibitem[Kazantzidis \etal(2004)]{Ka04}
Kazantzidis, S., Kravtsov, A. V., Zentner, A. R.; Allgood, B., 
Nagai, D., Moore, B. 2004 \apjl, 611, 73

\bibitem[Macci\`o \etal(2006)]{ma06}
Macci\`o, A., Moore, B., Stadel, J. 2006, \apjl, 636, L25

\bibitem[Kormendy \& Kennicutt(2004)]{kk04} 
Kormendy, J., Kennicutt, R.C. 2004, \araa, 42, 603

\bibitem[Krips \etal(2005)]{Kr05}
Krips, M., Eckart, A., Neri, R. \etal. 2005 \aap, 442, 479

\bibitem[Krips \etal(2007)]{Kr07}
Krips, M., Eckart, A., Krichbaum, T. P. \etal. 2007 \aap, 464, 553

\bibitem[Merrifield(2004)]{Me04}
Merrifield, M. 2004 IAU Symposium 220, 431

\bibitem[{Morganti \etal}(2006)]{Mo06}
Morganti, R., de Zeeuw, P. T., Oosterloo, T. A., 
McDermid, R. M., Krajnovic, D., Cappellari, M., Kenn, F., 
Weijmans, A., Sarzi, M. 2006 \mnras\ 371, 157

\bibitem[Nagar \etal(2005)]{Na05}
Nagar, N. M., Falcke, H., Wilson, A. S. 2005 \aap, 435, 521

\bibitem[{Noordermeer \etal}(2005)]{No05}
Noordermeer, E., van der Hulst, J. M., Sancisi, R., Swaters, R. A.,
van Albada, T. S. 2005 \aap\ 442, 137

\bibitem[Noordermeer(2006)]{No06}
Noordermeer, E. 2006, PhD Thesis, Groningen University, Netherlands.

\bibitem[{Noordermeer \etal}(2007)]{No07}
Noordermeer, E., van der Hulst, J. M., Sancisi, R., Swaters, R. A.
2007 \mnras\ 376, 1513


\bibitem[Peletier \& Willner(1993)]{PeWi93}
Peletier R. F., Willner S. P. 1993, \apj, 418, 626

\bibitem[Petric \& Rupen(2007)]{pr07}   
Petric, A. O., Rupen, M. P. 2007, \aj, 134, 1952 

\bibitem[Pott \etal(2004)]{Po04}
Pott, J.-U., Hartwich, M., Eckart, A., Leon, S., Krips, M.,
Straubmeier, C. 2004 \aap, 415, 27

\bibitem[Quillen \etal(1992)]{Q92}
Quillen, A. C., de Zeeuw, P. T., Phinney, E. S., Phillips, T. G. 1992
\apj, 391, 121

\bibitem[Quillen \etal(1993)]{Q93}
Quillen, A.C., Graham, J. R., Frogel, J. A. 1993 \apj, 412, 550

\bibitem[{Reshetnikov \& Combes}(1994)]{rc94}
Reshetnikov, V. P., Combes, F. 1994 \aap, 291, 57

\bibitem[Rice \etal(1988)]{Ri88}
Rice, W., Lonsdale, Carol J., Soifer, B. T., Neugebauer, G., Kopan, E. L., Lloyd, Lawrence A., de Jong, T., Habing, H. J. 1988 \apjs, 68, 91

\bibitem[Rix \& Zaritsky(1995)]{rz95} 
Rix, H.-W., Zaritsky, D. 1995, \apj, 447, 82

\bibitem[Roberts \& Haynes(1994)]{rh94}
Roberts, M. S., Haynes, M. P. 1994 \araa, 32, 115

\bibitem[Rots(1975)]{Ro75}
Rots, A. 1975, \aap, 45, 43

\bibitem[Ryden(2006)]{ry06} 
Ryden, B. S. 2006, \apj, 641, 773

\bibitem[Sackett \& Sparke(1990)]{SS90}
Sackett P. D., Sparke, L. S. 1990 \apj, 361, 408 

\bibitem[Sackett \etal(1994)]{S94}
Sackett P. D., Rix  H.-W., Jarvis B. J., Freeman K. C., 1994, \apj, 436, 629 

\bibitem[Sackett \& Pogge(1995)]{SP95}
Sackett P. D., Pogge R. W., 1995, in Dark Matter, AIP Conf.\ Ser.\
336, 141 

\bibitem[Sancisi \etal(2008)]{sa08}
Sancisi, R. Fraternali, F., Oosterloo, T., van der Hulst, J.M. 2008 \aapr, 15, 189

\bibitem[{Schaye}(2004)]{sch04}
Schaye, J. 2004 \apj\ 609, 667

\bibitem[Schwarz(1985)]{sc85}
Schwarz, U. J. 1985, \astap, 142, 273


\bibitem[Sparke(1986)]{Sp86}
Sparke L. S. 1986, \mnras, 219, 657

\bibitem[Sparke(1990)]{Sp90}
Sparke, L. S. 1990 in `Dynamics and Interactions of Galaxies',
ed. R. Wielen (Springer: Berlin), p338

\bibitem[Sparke(1996)]{Sp96}
Sparke L. S. 1996, \apj, 473, 810

\bibitem[Sparke(2002)]{Sp02}
Sparke L. S., 2002, in 
The Shapes of Galaxies and Their Dark Matter Halos,
ed P. Natarajan (World Scientific, Singapore), p178

\bibitem[Sparke(2004)]{Sp04}
Sparke, L. S. in High Velocity Clouds, eds. H. van Woerden, 
B. P. Wakker, U. J. Schwarz \& K. S. de Boer, Ap. Sp. Sci.
Library 312, 273 (Kluwer Academic Publishers, Dordrecht, Netherlands)

\bibitem[Sparke \etal(2008)]{Sp08}
Sparke, L. S., van Moorsel, G., Erwin, P., Wehner, E. M. H. 2008, \aj, 135, 99

\bibitem[Steiman-Cameron \& Durisen(1984)]{SCD84}
Steiman-Cameron, T. Y., Durisen, R. H. 1984 \apj, 276, 101

\bibitem[Steiman-Cameron \& Durisen(1988)]{SCD88}
Steiman-Cameron, T. Y., Durisen, R. H. 1988 \apj, 325, 26

\bibitem[Steiman-Cameron \etal(1992)]{SCKD92}
Steiman-Cameron, T. Y., Kormendy, J., Durisen, R. H. 1992 \aj,  104,
1339

\bibitem[Swaters \& Rubin(2003)]{SR03}
Swaters, R. A., Rubin, V. C. 2003, \apj, 587, L23


\bibitem[Trinh \etal(2006)]{Tr07}
Trinh, Christopher Q., Sparke, L. S., Gallagher, J. S. 2006 BAAS 209.1807

\bibitem[Tully \& Fouqu\'e (1985)]{TF85}
Tully, R. B., Fouqu\'e, P. 1985 \apjs, 58, 67

\bibitem[Tully(1998)]{Tu98}
Tully, R. B. 1988, Nearby Galaxies Catalog, Cambridge University Press

\bibitem[Tully \etal(1996)]{Tu96}
Tully, R. B., Verheijen, M. A. W., Pierce, M. J., Huang,
J.-S. \& Wainscoat, R. J. 1996 \aj, 112, 2471

\bibitem[Tully \& Pierce(2000)]{TuPi00}
Tully  R. B., Pierce, M.J. 2000 \apj, 533, 744

\bibitem[{van Albada \etal}(1982)]{vA82}
van Albada, T. S., Kotanyi, C. G., Schwarzschild, M. 1982, \mnras,
198, 303

\bibitem[van der Hulst \& Sancisi(2005)]{vdhs05} 
 van der Hulst, J. M., Sancisi, R. 2005, ASP Conference Series, 331, 139  

\bibitem[{van~Driel \etal}(1995)]{vD95}
van~Driel, W., Combes, F., Casoli, F. \etal, 1995 \aj, 109, 942

\bibitem[{van~Driel \etal}(2000)]{vD00}
van~Driel, W., Arnaboldi, M., Combes, F., \&\ Sparke, L. S. 2000, \aaps,
141, 385

\bibitem[{van Gorkom \etal}(1987)]{vGKS}
van Gorkom, J. H., Schechter, P. L., Kristian, J. 1987, \apj, 314, 457

\bibitem[{Verheijen \& Sancisi}(2001)]{VeSa01}	
Verheijen, M. A. W., Sancisi, R., 2001, \aap, 370, 765

\bibitem[{Verheijen}(2001)]{Ve01}
Verheijen, M. A. W. 2001, \apj, 563, 694

\bibitem[{Vogelaar \& Terlouw}(2001)]{vt01}
Vogelaar, M. G. R., Terlouw, J. P. 2001 in 
Astronomical Data Analysis Software and Systems X, ASPC 238, 358;
also http://www.astro.rug.nl/[tilde]gipsy/

\bibitem[{Whitmore \etal}(1990)]{PRC}
Whitmore, B.~C., Lucas, R.~A., McElroy, D.B., Steiman-Cameron, T.~Y.,
Sackett, P.~D., \&\ Olling, R.~P. 1990, \aj, 100, 1489 

\bibitem[Whitmore(1984)]{Wh84}
Whitmore, B.C. 1984, \aj, 89, 618

\end{thebibliography}
\end{document}